\documentclass[12pt,preprint,preprint,nofootinbib]{revtex4}
\pdfoutput=1
\usepackage{epsf, array, color}
\usepackage{graphicx}
\usepackage{subfigure}
\usepackage{bbold}
\usepackage{amsmath}
\usepackage[linktocpage,colorlinks,urlcolor=blue]{hyperref}
\usepackage{mathtools}
\usepackage{dcolumn}
\usepackage{url}
\usepackage{verbatim}
\usepackage{wrapfig}
\usepackage{amsfonts}
\usepackage{comment}
\usepackage{epigraph}
\usepackage{slashed}
\usepackage[utf8]{inputenc}

\usepackage{bbm}
\usepackage{cancel}
\usepackage{epsf, array, color}

\newcommand{\be}{\begin{equation}}
\newcommand{\ee}{\end{equation}}

\newcommand{\bea}{\begin{eqnarray}}
\newcommand{\eea}{\end{eqnarray}}
\newcommand{\nn}{\nonumber \\}

\def\beq{\begin{equation}}
\def\eeq{\end{equation}}

\newcommand{\co}{\mathcal{O}}

\newcommand{\C}{\mathcal{C}}

\def\sq2{\sqrt{2}}
\newcommand{\nc}{\newcommand}

\newcommand{\smallh}{{\scriptscriptstyle H}}
\newcommand{\mh}{m_\smallh}

\newcommand{\Op}{{\cal O}}

\renewcommand{\arraystretch}{1.2}
\nc{\btb}{\begin{tabular}}    
\nc{\etb}{\end{tabular}}

\nc{\vp}{\phi}
\nc{\tvp}{\widetilde{\phi}}
\nc{\vpj }{\mbox{${\vp^\dag i\,\raisebox{2mm}{\boldmath ${}^\leftrightarrow$}\hspace{-4mm} D_\mu\,\vp}$}}
\nc{\vpjt}{\mbox{${\vp^\dag i\,\raisebox{2mm}{\boldmath ${}^\leftrightarrow$}\hspace{-4mm} D_\mu^{\,I}\,\vp}$}}

\def\mhsm{M_H}
\def\qq2{q^2}
\def\qq4{q^4}
\def\cphib{{\mathcal {C}}_{\phi B}}
\def\cphiw{{\mathcal {C}}_{\phi W}}
\def\cphiwb{{\mathcal{C}}_{\phi WB}}
\def\cll{{\mathcal{C}}^{ll}_{\mu e e \mu}}
\def\chl3{{\mathcal{C}}_{\phi l}^{(3)}}
\def\cphid{{\mathcal{C}}_{\phi D}}
\def\cphik{{\mathcal{C}}_{\phi\square}}
\def\nn{\nonumber}
\def\gev{\rm GeV}
\def\tev{\rm TeV}

\def\h{H} 
\def\mh2{M_H^2}
\def\mh4{M_H^4}
\def\mz{M_Z}
\def\mw{M_W}
\def\hoz{{M_H^2\over M_Z^2}}
\def\thz{{4M_t^2-M_H^2\over M_Z^2}}

\def\g1bar{{\overline g}_1}
\def\g2bar{{\overline g}_2}
\def\gf{G_\mu}

\def\g1{g_1}
\def\g2{g_2}

\begin{document}

\title{Higgs Decays  to $ZZ$
and $Z\gamma$  in the SMEFT:\\ an NLO analysis}

\author{S.~Dawson and P.P. Giardino  }
\affiliation{
\vspace*{.5cm}
  \mbox{
  Department of Physics,\\
  Brookhaven National Laboratory, Upton, N.Y., 11973,  U.S.A.}\\
 \vspace*{1cm}}

\begin{abstract}
We calculate the complete  one-loop electroweak corrections to the inclusive $\h\rightarrow ZZ$ and
$\h\rightarrow Z\gamma$ decays in the dimension-$6$ extension of the Standard Model
Effective Field Theory (SMEFT).  The corrections to $\h\rightarrow ZZ$ are computed for on-shell $Z$ bosons and are 
a precursor to the physical $\h\rightarrow Z f {\overline{f}}$ calculation.  We present compact numerical formulas
for our results and demonstrate that the logarithmic contributions that result from the renormalization group evolution
of the SMEFT coefficients are larger than the finite NLO contributions to the decay widths.  
As a by-product of our calculation, we obtain the first complete result for the finite corrections to 
$G_\mu$ in the SMEFT.
\end{abstract}

\maketitle

\section{Introduction}

The LHC experimental discovery of the Higgs boson, along with
the measurement of Higgs properties that are in rough agreement with the Standard Model (SM) predictions, suggests that
the SM is a valid effective theory at the weak scale.  The lack of new particles up to the $\tev$ scale makes
possible the parameterization of possible high scale 
physics  effects in terms of higher dimension operators containing only SM fields \cite{Buchmuller:1985jz}. In this paper we study new physics 
contributions to Higgs decays in the context of the dimension-$6$ Standard Model Effective Field Theory (SMEFT).
When compared with precise theoretical calculations, 
measurements
of Higgs properties serve to constrain the coefficients of the higher dimension operators and restrict possible
beyond the SM (BSM) physics at energies $\Lambda>>v$.

We study Higgs decays to $Z$ boson pairs and to $Z\gamma$ in the context of the SMEFT,  where  new physics is described
by a tower of operators,
\begin{equation}
{\cal L}={\cal L}_{SM}+\Sigma_{k=5}^{\infty}\Sigma_{i=1}^n {\mathcal{C}_i^k\over \Lambda^{k-4}} O_i^k\, .
\label{eq:lsmeft}
\end{equation}
The dimension-$k$ operators are constructed from SM fields and the
 BSM physics effects reside in the coefficient functions, $C_i^k$.  
For large $\Lambda$,  it is sufficient to retain only the lowest dimensional operators.
We assume lepton number conservation, so the lowest dimension relevant operators are dimension-$6$.
Ignoring flavor, there are $59$ dimension-$6$ operators that 
are $SU(3)\times SU(2)\times U(1)$ invariant combinations of the SM fields \cite{Buchmuller:1985jz,Grzadkowski:2010es}.  
The operators have been classified in several different bases, which 
are related by the equations of motion \cite{Buchmuller:1985jz,Grzadkowski:2010es,Hagiwara:1993ck,Giudice:2007fh}.  
In this paper we will use the Warsaw basis of 
Ref. \cite{Grzadkowski:2010es}. 

A detailed understanding of Higgs properties requires the inclusion of the  dimension-$6$ tree level SMEFT effects, 
along with radiative corrections in the effective field theory. Measurements
at the Higgs mass scale, $\mhsm,$  that are sensitive to a set of SMEFT coefficients, $\C_i(\mhsm$), at leading order will develop
logarithmic sensitivity to other coefficients when renormalization group
evolved to the scale $\Lambda$, due to the renormalization group mixing of the 
coefficients \cite{Elias-Miro:2013mua,Elias-Miro:2013gya,Jenkins:2013zja,Jenkins:2013wua,Alonso:2013hga}. There are also finite contributions 
 that may be of the same numerical size as the logarithmic terms.  

We compute the one-loop electroweak SMEFT contributions to the decays $\h\rightarrow ZZ$ and $\h\rightarrow Z\gamma$. 
 These corrections include
the one-loop SM electroweak corrections, along with the one-loop corrections due to the SMEFT operators of Eq.~(\ref{eq:lsmeft}).
Our results are interesting from a purely theoretical perspective and we present the first complete one-loop SMEFT renormalization
of $G_\mu$.  The one-loop SMEFT corrections to $
\h\rightarrow b {\overline b}$ \cite{Gauld:2015lmb,Gauld:2016kuu} and 
$\h\rightarrow \gamma\gamma$ \cite{Hartmann:2015oia, Hartmann:2015aia}
are known, as well as a general NLO SMEFT calculation of $2$-body Higgs decays \cite{Ghezzi:2015vva}.
 The physical process for 
$\mhsm=125~\tev$
is $\h\rightarrow Z f {\overline f}$ and our calculation is a precursor to the eventual one-loop $3$-body SMEFT calculation. 

We review the SMEFT framework in Sec. \ref{sec:model} and our renormalization framework in Sec. \ref{sec:renorm}.  Sec. \ref{sec:hh}
contains numerical results for $\h\rightarrow ZZ$ and $\h\rightarrow Z\gamma$ and a discussion of the phenomenological impact of
our calculation.  A comparison with the physical off-shell process, $\h\rightarrow Z f {\overline{f}}$,  is in Appendix \ref{sec:appa}.  Appendix
\ref{sec:appfit} contains numerical fits for the one-loop SMEFT result for $\h\rightarrow ZZ$ and 
Appendix \ref{sec:appf} has analytic formulas for the logarithmic contributions
to the one-loop SMEFT result for $\h\rightarrow ZZ$. Lastly, Appendix \ref{sec:appdelta} contains  the one-loop SMEFT calculation of $G_\mu$.
 
\section{smeft Basics}
\label{sec:model}
In this section we briefly introduce the SMEFT.  We consider the Lagrangian in Eq.~(\ref{eq:lsmeft}) truncated at dimension$-6$,
\beq
\mathcal{L}=\mathcal{L}_{\text{SM}}+\mathcal{L}^{(6)}_{\text{EFT}};~~~~\mathcal{L}^{(6)}_{\text{EFT}}=\sum_i 
{\mathcal{C}^{(6)}_i\over\Lambda^2}  \mathcal{O}^{(6)}_i,
\eeq
where $\mathcal{L}_{\text{SM}}$ is the Standard Model Lagrangian, and $\mathcal{L}^{(6)}_{\text{EFT}}$ is the most general 
$SU(3)\times SU(2)\times U(1)$ invariant EFT Lagrangian that can be built using only dimension-$6$ operators. In the following we  drop the superscript $(6)$. 
Only a few operators contribute to the $H\to ZZ$ and $H\to Z\gamma$ decays at tree-level, while more  operators
contribute  at one- loop. In total, 19 of the 59 independent dimension-$6$ operators of the Warsaw basis enter our calculation,
\begin{eqnarray}
&&\mathcal{O}_W, \mathcal{O}_\phi, \mathcal{O}_{\phi\square}, \mathcal{O}_{\phi D}, \mathcal{O}_{u\phi}, \mathcal{O}_{\phi W}, \mathcal{O}_{\phi B},  \mathcal{O}_{\phi WB}, \mathcal{O}_{uW}, 
\nonumber \\ &&
\mathcal{O}_{uB}, \mathcal{O}_{\phi l}^{(1)}, \mathcal{O}_{\phi l}^{(3)}, \mathcal{O}_{\phi e}, \mathcal{O}_{\phi q}^{(1)}, \mathcal{O}_{\phi q}^{(3)}, \mathcal{O}_{\phi u}, \mathcal{O}_{\phi d}, \mathcal{O}_{ll}, \mathcal{O}_{lq}^{(3)}\, ,
\label{eq:Operators}
\end{eqnarray}
where the operators are defined in Tab. \ref{tab:opdef},
\begin{equation}
D_\mu=\partial_\mu+ig^\prime B_\mu Y+ig{\tau^I\over 2}W_\mu^I +ig_s T^AG_\mu^A\, ,
\end{equation}
 $\tau^I$ are the Pauli matrices and $l_L$ and $q_L$ are the $SU(2)_L$ doublet  lepton and quark fields. 
For simplicity, we assume a diagonal flavor structure for the coefficients $\mathcal{C}$, {\it i.e.} $\underset{p,r}{\C_i}=\C_i \underset{p,r}{\mathbb{1}}$,
where $p,r$ are flavor indices. 
 Furthermore, we assume $\underset{e,\mu,\mu,e}{\C_{ll}}=\underset{\mu,e,e\mu}{\C_{ll}}\equiv \C_{ll}$ and 
 $\underset{\mu,\mu,t,t}{\C_{lq}^{(3)}}=\underset{e,e,t,t}{\C_{lq}^{(3)}}\equiv \C_{lq}^{(3)}$.

In general, the presence of the dimension-$6$ operators changes the structure of the Lagrangian and the correlations between
the  Lagrangian quantities and  the measured observables \cite{Falkowski:2015fla,Brivio:2017bnu}. In the following we discuss these modifications, as they are relevant to the one-loop SMEFT calculations
of $\h\rightarrow ZZ$ and $\h\rightarrow Z\gamma$.

\begin{table}[t] 
\centering
\renewcommand{\arraystretch}{1.5}s
\begin{tabular}{||c|c||c|c||c|c||} 
\hline \hline
$\Op_W$                & $\epsilon^{IJK} W_\mu^{I\nu} W_\nu^{J\rho} W_\rho^{K\mu}$ &    
 $\Op_\vp$       & $(\vp^\dag\vp)^3$ &
 $\Op_{\vp\Box}$ & $(\vp^\dag \vp)\raisebox{-.5mm}{$\Box$}(\vp^\dag \vp)$ 
 \\
$\Op_{\vp D}$   & $\left(\vp^\dag D^\mu\vp\right)^* \left(\vp^\dag D_\mu\vp\right)$ &
$\underset{p,r}{\Op_{u\vp}}$           & $(\vp^\dag \vp)(\bar q'_p u'_r \tvp)$&
 $\Op_{\vp W}$    & $ (\vp^\dag \vp)\, W_{\mu\nu} W^{\mu\nu}$ 
   \\
   $\Op_{\vp B}$     & $ (\vp^\dag \vp)\, B_{\mu\nu} B^{\mu\nu}$ &
     $\Op_{\vp WB}$     & $ (\vp^\dag \tau^I \vp)\, W^I_{\mu\nu} B^{\mu\nu}$ &
$\Op_{uW}$               & $(\bar q'_p \sigma^{\mu\nu} u'_r) \tau^I \tvp\, W_{\mu\nu}^I$ 
\\
$\underset{p,r}{\Op_{uB}}$        & $(\bar q'_p \sigma^{\mu\nu} u'_r) \tvp\, B_{\mu\nu}$&
$\underset{p,r}{\Op_{\vp l}^{(1)}}$      & $(\vpj)(\bar l'_p \gamma^\mu l'_r)$ &
$\underset{p,r}{\Op_{\vp l}^{(3)}}$      & $(\vpjt)(\bar l'_p \tau^I \gamma^\mu l'_r)$
\\
 $\underset{p,r}{\Op_{\vp e}}$            & $(\vpj)(\bar e'_p \gamma^\mu e'_r)$& 
$\underset{p,r}{\Op_{\vp q}^{(1)}}$      & $(\vpj)(\bar q'_p \gamma^\mu q'_r)$&
$\underset{p,r}{\Op_{\vp q}^{(3)}}$      & $(\vpjt)(\bar q'_p \tau^I \gamma^\mu q'_r)$
\\
$\underset{p,r}{\Op_{\vp u}}$            & $(\vpj)(\bar u'_p \gamma^\mu u'_r)$  &
 $\underset{p,r}{\Op_{\vp d}}$            & $(\vpj)(\bar d'_p \gamma^\mu d'_r)$ &
    $\underset{p,r,s,t}{\Op_{ll}}$        & $(\bar l'_p \gamma_\mu l'_r)(\bar l'_s \gamma^\mu l'_t)$  
    \\
 $\underset{p,r,s,t}{\Op_{lq}^{(3)}}$                & $(\bar l'_p \gamma_\mu \tau^I l'_r)(\bar q'_s \gamma^\mu \tau^I q'_t)$  
  &&
  &&
\\
\hline \hline
\end{tabular}
\caption{Dimension-6 operators relevant for our study
  (from \cite{Grzadkowski:2010es}). For brevity we suppress fermion
  chiral indices $L,R$. $I=1,2,3$ is an $SU(2)$ index, $p,r$ are flavor
  indices,  and $\vpj\equiv \phi^\dagger D_\mu \phi- (D_\mu\phi^\dagger) \phi$.  \label{tab:opdef}}
\end{table}

\subsection{Higgs sector}
In the SMEFT, the Higgs Lagrangian takes the form,
\bea
\mathcal{L}&=&(D^\mu \phi )^\dagger (D_\mu \phi )+ \mu ^2 \phi ^\dagger \phi  -\lambda (\phi ^\dagger \phi )^2 \nn\\
&+& \frac1{\Lambda^2} \biggl(
\C_\phi  (\phi ^\dagger \phi )^3+\C_{\phi \square}  (\phi ^\dagger \phi )\square(\phi ^\dagger \phi )
+\C_{\phi D}(\phi ^\dagger D^\mu \phi)^*(\phi^\dagger D_\mu\phi)\biggr)
\label{eq:HiggsPotential}
\eea
where  $\phi$  is the Higgs doublet:
\bea
\phi =\left(
\begin{array}{c}
\phi^+ \\
\frac1{\sqrt{2}}(v+H+i \phi^0)
\end{array}
\right),
\eea
and $v$ is the vacuum expectation value (vev) defined as the minimum of the potential, 
\beq
v\equiv\sqrt{2}\langle \phi \rangle= \sqrt\frac{\mu^2}{\lambda}+\frac{3 \mu^3}{8 \lambda^{5/2}}\frac{\C_\phi }{\Lambda^2}.
\eeq
Due to the presence of $\co_{\phi \square}$ and $\co_{\phi D}$ in Eq.~(\ref{eq:HiggsPotential}), the kinetic terms in the resulting Lagrangian are not canonically normalized.
As a consequence we need to shift the fields, 
\bea
\h&\to& \h\biggl(1-\frac{v^2}{\Lambda^2}(\frac14 \C_{\phi D}-\C_{\phi \square})\biggr), \nn\\
\phi^0&\to&\phi^0\biggl(1-\frac{v^2}{\Lambda^2}(\frac14 \C_{\phi D})\biggr), \nn\\
\phi^+&\to&\phi^+, 
\label{eq:HiggsShift}
\eea
and the physical mass of the Higgs, defined as the pole of the propagator, becomes,
\beq
\mhsm^2=2\lambda v^2-\frac{v^4}{\Lambda^2}(3 \C_\phi -4 \lambda \C_{\phi \square}+\lambda C_{\phi D}).
\eeq
As anticipated,  the relation between the Lagrangian parameters and the measured observable  ($\mhsm$) is altered by the presence of the dimension-$6$ operators \cite{Falkowski:2015fla,Brivio:2017bnu}.

\subsection{Gauge sector}
The introduction of the operators in Eq.~(\ref{eq:Operators})  also alters the form of the kinetic terms  for the Lagrangian of the gauge sector. 
The relevant Lagrangian terms are:
\bea
\mathcal{L}&=&-\frac14 W{I,\mu\nu} W_{\mu\nu}^I-\frac14 B^{\mu\nu}B_{\mu\nu}\nn\\
&+&\frac1{\Lambda^2}\bigg(\C_{\phi W} (\phi ^\dagger \phi )W^{I,\mu\nu} W_{\mu\nu}^I+\C_{\phi B} (\phi ^\dagger \phi )B^{\mu\nu} B_{\mu\nu}+\C_{\phi WB} (\phi ^\dagger \tau^I \phi )W^{I,\mu\nu} B_{\mu\nu} \bigg),
\eea
It is  convenient to define "barred" fields, ${\overline W}_\mu\equiv (1-\C_{\phi W} v^2/\Lambda^2)W_\mu$ and ${\overline B}_\mu\equiv (1-\C_{\phi B}v^2/\Lambda^2)B_\mu$ and "barred" gauge couplings,  ${\overline g}_2\equiv (1+\C_{\phi W} v^2/\Lambda^2)g_2$ and ${\overline g}_1\equiv (1+\C_{\phi B}v^2/\Lambda^2)g_1$ so that ${\overline W}_\mu {\overline g}_2= W_\mu g_2$ and ${\overline B}_\mu {\overline g}_1= B_\mu g_1$. The "barred" fields defined in this way have their kinetic terms properly normalized and preserve the form of the covariant derivative.  
The masses of the W and Z fields (poles of the propagators) are then expressed in 
terms of the "barred" couplings \cite{Dedes:2017zog,Alonso:2013hga}:
\bea
M_W^2&=&\frac{{\overline g}_2^2 v^2}4,\nn\\
M_Z^2&=&\frac{({\overline g}_1^2+{\overline g}_2^2) v^2}4+\frac{v^4}{\Lambda^2}\left(\frac18 ({\overline g}_1^2+{\overline g}_2^2) \C_{\phi D}+\frac12 {\overline g}_1{\overline g}_2\C_{\phi WB} \right).
\eea
It is interesting to note that the extra terms in the definition of the $Z$ mass are due to the rotation, $(W_\mu^3,B_\mu)\to(Z_\mu,A_\mu)$,
 that is proportional to $\C_{\phi WB}$ and the shift of $\phi^0$ in Eq.~(\ref{eq:HiggsShift}) that is proportional to $\C_{\phi D}$.

\subsection{Fermion sector}
Lastly, we study the fermion sector. We notice that the presence of the dimension-6 operators do not alter the kinetic terms in the Lagrangian, 
so we concentrate  on the mass terms\footnote{We neglect flavor mixing, so $\C_{u\phi}$ represents generically $\C_{u\phi}$, $\C_{c\phi}$, $\C_{t\phi}$, etc.}: 
\bea
\mathcal{L}&=&-(y_e \bar{l}_L e_R \phi+y_u \bar{q}_L u_R \tilde{\phi }+y_d \bar{q}_L d_R \phi +h.c.)\nn\\
&+& \frac1{\Lambda^2}\left( \C_{e\phi}(\phi ^\dagger \phi )( \bar{l}_L e_R \phi) + \C_{u\phi}(\phi ^\dagger \phi )(\bar{q}_L u_R \tilde{\phi })   + \C_{d\phi}(\phi ^\dagger \phi )(\bar{q}_L d_R \phi)  +h.c.   \right).
\label{eq:masses} 
\eea
The masses of all fermions are shifted by the interactions of Eq.~(\ref{eq:masses}).  The lepton and light quark masses do not enter the $1$-loop result for $\h\rightarrow ZZ$ and
can be safely set to $0$ there.  
However, the masses of the leptons and lighter quarks contribute logarithmically to the  $\gamma \gamma $ wave-function in the one-loop  $\h\to Z\gamma$  calculation and we
retain finite fermion masses there.  However, since  the lowest order (LO)
 $\h \to Z\gamma$  amplitude is  ${\cal {O}}({ v^2\over \Lambda^2})$, the contributions of the terms proportional to $\C_{e\phi}$ and $\C_{d\phi}$ to 
 light fermion masses, $m_f$, are ${\cal {O}}({m_fv^4\over \Lambda^4})$ and 
can be neglected. 
We concentrate  on the definition of the top pole mass, 
\beq
M_t=\frac{v}{\sqrt{2}}(y_t-\frac12\C_{u\phi}\frac{v^2}{\Lambda^2}).
\eeq

Dimension-6 operators involving fermions also  give contributions to the decay of the $\mu$ lepton, thus changing the relation between the 
vev, $v$, and the Fermi constant $G_\mu$ obtained from the measurement of the $\mu$ lifetime. Considering only contributions that interfere with the
 SM amplitude,  we obtain the  tree level result,
\begin{eqnarray}
G_\mu& =&\frac1{\sqrt{2} v^2}-\frac1{2\sqrt{2}\Lambda^2}
(\underset{e,\mu,\mu,e}{\C_{ll}}+\underset{\mu,e,e\mu}{\C_{ll}})+{\sqrt{2}\over 2\Lambda^2}
(\underset{e,e}{\C_{\phi l}}^{(3)}+\underset{\mu,\mu}{\C_{\phi l}}^{(3)})
\nonumber \\
&\equiv &\frac1{\sqrt{2} v^2}-\frac1{\sqrt{2}\Lambda^2}\C_{ll}+{\sqrt{2}\over \Lambda^2}\C_{\phi l}^{(3)},
\end{eqnarray}
where in the last equality we assumed flavor universality of the coefficients.

\section{Renormalization}
\label{sec:renorm}
The SM one-loop electroweak corrections to $H\to ZZ$
are well known and we reproduce the results of Ref. \cite{Kniehl:1990mq}.    The decay $H\to Z\gamma$  first occurs at one-loop in
the SM and analytic results are in Ref. \cite{Cahn:1978nz,Bergstrom:1985hp}. 

The calculations of the radiative corrections to $H\to ZZ$ and $H\to Z\gamma$ in the SMEFT proceed in the usual way \cite{Denner:1991kt}
 by the choice of a renormalization scheme, the definition of a suitable set of input parameters and the calculation of the 1PI amplitudes involved.
However, since the SMEFT theory is only renormalizable order by order in the $(v^2/\Lambda^2)$ expansion, 
we need to add an extra requirement and drop all  terms proportional to $(v^2/\Lambda^2)^a$ with $a>1$.
These terms would need counterterms of dimension-$8$ that are not included in our study. Dropping them makes it possible for us to proceed with our renormalization program. 
The one-loop SMEFT calculation contains  both tree level
and one-loop
contributions from the dimension-6 operators, along with the  full electroweak one-loop SM amplitude.
We chose a modified on shell (OS) scheme, where the SM parameters are OS quantities, while the SMEFT 
coefficients are defined as ${\overline{MS}}$ quantities.

 The tadpole counterterms are defined such that they cancel completely the tadpole graphs \cite{Fleischer:1980ub}. 
 This condition forces us to identify the renormalized vacuum to be the minimum of the renormalized scalar potential
 at each order of perturbation theory. 
 Notice that, due to this choice, the intermediate quantities defined here are gauge dependent, while the final result is gauge independent as expected.

We choose the $G_\mu$ scheme, where we take the physical  input parameters to be\footnote{The light quark masses and lepton masses enter
into the $\gamma$ wave-function renormalization for $\h\rightarrow Z\gamma$. We take $m_\tau=1.776~\gev,\, m_\mu=0.105~\gev\,\text{and}\, m_e=0.0005~\gev$ for the lepton masses, while the light quark contribution has been replaced with the hadronic contribution to the vacuum polarization $\Delta\alpha_{\rm had}^{(5)}=0.02764 \pm 0.00007 $\cite{Tanabashi:2018oca}.},
\begin{eqnarray}
G_\mu&=&1.1663787(6)\times 10^{-5} \gev^{-2}\nonumber \\
M_Z&=&91.1876\pm .0021\gev\nonumber \\
M_W&=&80.385\pm .015~\gev\nn\\
\mhsm&=&125.09\pm 0.21\pm 0.11 ~\gev\nn\\
M_t&=&173.1\pm0.6~\gev\nn\, .
\end{eqnarray}

Since the coefficients of the dimension-6 operators are not measured quantities, it is convenient to treat them as $\overline{{MS}}$
parameters, so the renormalized coefficients are  $\C(\mu)=\C_0-\text{poles}$, where $\C_0$ are the bare quantities. 
The poles of the coefficients $\C_0$ are obtained from the renormalization group evolution of the coefficients computed in the unbroken
phase of the theory in 
Refs. \cite{Jenkins:2013zja,Jenkins:2013wua,Alonso:2013hga}. In general, one can write,
\beq
\C_i(\mu)=\C_{0,i}-\frac1{2\hat{\epsilon}}\frac1{16\pi^2}\gamma_{ij}\C_j,
\eeq
where $\mu$ is the renormalization scale,  $\gamma_{ij}$ is the one-loop anomalous dimension, 
\beq
\mu \frac{d \C_i}{d\mu}=\frac1{16\pi^2}\gamma_{ij}\C_j,
\eeq
and $\hat{\epsilon}^{-1}\equiv\epsilon^{-1}-\gamma_E+\log(4\pi)$ is related to the regulator $\epsilon$ for integrals evaluated in $d=4-2\epsilon$ dimensions.

At one-loop, the tree level parameters of the previous section (denoted with the subscript 0 in this section) must be renormalized.
The renormalized SM masses are defined by,
\beq
M_V^2=M^2_{0,V}-\Pi_{V}(M^2_{V}),
\eeq
where $\Pi_{V}(M^2_V)$ is the one-loop correction to the 2-point function of either Z or W computed on-shell.
The gauge boson $2$- point functions in the SMEFT can be found in Refs. \cite{Chen:2013kfa,Ghezzi:2015vva}.

The one- loop relation between the vev and the Fermi constant is defined by the equation,
\beq
G_\mu+{\C_{ll}\over\sqrt{2}\Lambda^2}-\sqrt{2}{\C_{\phi l}^{(3)}\over\Lambda^2}\equiv\frac1{\sqrt{2} v_0^2}(1+\Delta r),
\label{eq:geftdef}
\eeq
where $v_0$ is the unrenormalized minimum of the potential and
 $\Delta r$ is obtained from the one-loop corrections to $\mu$ decay and is given by
\begin{eqnarray}
\Delta r&=&2 v^2 \mathcal{B} + \mathcal{V}(1+\frac{v^2}{\Lambda^2}\C_{\phi l}^{(3)}) + \mathcal{E}(1+\frac{v^2}{\Lambda^2}(2\,\C_{\phi l}^{(3)}-\C_{ll}))
\nonumber \\ &&  - \frac{A_{WW}}{M_W^2}(1+2\frac{v^2}{\Lambda^2}\C_{\phi l}^{(3)})+\frac1{16\pi^2}\frac1{2\hat{\epsilon}}\frac{v^2}{\Lambda^2}(2\gamma_{\phi l,j}^{(3)}\C_j-\gamma_{ll,j}\C_j).
\label{eq:deltar}
\end{eqnarray}
In Eq. (\ref{eq:deltar}), $\mathcal{B}$ is the box contribution, $\mathcal{V}$ is the vertex contribution, $A_{WW}=\Pi_{W}(0)$ is the W boson self-energy at zero momentum and $\mathcal{E}$ is the sum of the lepton ($\mu,\, e,\, \nu_\mu,\, \nu_e$) wave-function renormalizations. All the quantities are calculated at zero external momenta. Notice that the definition of $\Delta r$ is modified with respect to the SM result 
due to the presence of dimension-6 operators in the tree-level relation between $G_\mu$ and $v$ given in Eq. (\ref{eq:geftdef}).  Additionally,
 we absorb the poles of the coefficients $\C$ into the definition of $\Delta r$. The renormalization of the vev is then,
\begin{eqnarray}
v^2&=&v_0^2-\delta v^2\nonumber \\
   \delta v^2&=&\frac{\Delta r}{\sqrt{2}G_\mu}
   \biggl(1-\frac1{\sqrt{2}G_\mu\Lambda^2}\C_{ll}+\frac{\sqrt{2}}{G_\mu\Lambda^2}\C_{\phi l}^{(3)}\biggr)\, .
\end{eqnarray}
Analytic expressions for $\Delta r$ in both the SM and the SMEFT at dimension-$6$ are given in Appendix C in
the $R_{\xi}$. 

In the following, we indicate with the symbol $\Delta \mathcal{A}^{\mu\nu}$ the sum of the contributions
from the renormalization of the vev, the masses, and the SMEFT coefficients  described above.
  The other contributions to the one-loop corrections are the proper one-particle irreducible amplitudes $\mathcal{A}_{1PI}^{\mu\nu}$, 
  the particle reducible contributions $\mathcal{A}_{PR}^{\mu\nu}$ due to the $Z/\gamma$ mixing which
  arises in the $\h\rightarrow ZZ$ process,
   and the external leg wave-function renormalization $\delta Z_i=-\partial \Pi_i(k^2)/\partial k^2 |_{M_i^2}$.  The calculation of these contributions is relatively straightforward. 
  We start with the $R_\xi$ Feynman rules for the SMEFT in the Warsaw basis presented in \cite{Dedes:2017zog} and convert them to  a FeynArts  \cite{Hahn:2000kx}
  model file, using the FeynRules \cite{Alloul:2013bka} routines, from which we obtain
   the amplitudes needed for our calculation. We reduce the integrals in terms of Passarino-Veltman integrals \cite{Passarino:1978jh}, using FeynCalc \cite{Mertig:1990an,Shtabovenko:2016sxi}
   and lastly we use LoopTools \cite{Hahn:2000jm} to  numerically calculate the integrals.
Once we compute all the terms that contribute, the one-loop correction can be simply written as
\beq
\mathcal{A}^{1l,\mu\nu}=\mathcal{A}_{1PI}^{\mu\nu}
+\mathcal{A}^{0l,\mu\nu} \frac12 \sum_i \delta Z_i + \mathcal{A}_{PR}^{\mu\nu} + \Delta \mathcal{A}^{\mu\nu}.
\eeq
We verified that $\mathcal{A}^{1l,\mu\nu}$ is UV and IR finite and  we confirmed
its gauge invariance by computing the amplitudes  for the $H\to ZZ$ and $H\to Z\gamma$ processes in $R_{\xi}$ gauge.

We conclude this section with a 
few remarks on the truncation of the expansion in loops and powers of $\frac{v^2}{\Lambda^2}$. 
To clarify our explanation we consider the generic form of the  $H\rightarrow ZZ$ amplitude where we re-introduce the notation $\C^{(6)}$ and $\C^{(8)}$ for the coefficients of dimensions 6 and 8 operators:
\beq
\mathcal{A}\sim{\hat a}_0 g_{SM} + {\hat a}_1 \C^{(6)}\frac{v^2}{\Lambda^2} 
+ {\hat a}_2 \frac{g^3_{SM}}{16 \pi^2}+{\hat a}_3 \C^{(6)}\frac{v^2}{\Lambda^2}\frac{g^2_{SM}}{16 \pi^2}
+\frac{v^4}{\Lambda^4} ({\hat a}_4 (\C^{(6)})^2\frac{g_{SM}}{16 \pi^2}+ {\hat a}_5 \C^{(8)})+\dots
\label{eq:expA}
\eeq
In  Eq.~(\ref{eq:expA}) we assumed the following ordering $g_{SM}>\C^{(6)} \frac{v^2}{\Lambda^2}>\frac{g^3_{SM}}{16\pi^2}>\C^{(8)}\sim (\C^{(6)})^2\frac{g_{SM}}{16 \pi^2}$, and we ordered the terms from larger to smaller according to it.
Note that in Eq.~(\ref{eq:expA}), we have only included one insertion of dimension-6 operators at tree level, $ ({\hat{a}}_1)$.  If we were considering a more complicated process, it would in general be possible to have two (or more) tree level insertions of dimension-6 operators, which would give a contribution of ${\cal{O}}({v^4\over \Lambda^4})$ that would be of the same order as terms we have included.
The squared amplitude then is, 
\bea
|\mathcal{A}|^2&\sim&{\hat a}_0^2 g^2_{SM} +2 
{\hat a}_0 {\hat a}_1 \C^{(6)}\frac{v^2}{\Lambda^2} g_{SM} +
{\hat  a}_1^2 (\C^{(6)})^2\frac{v^4}{\Lambda^4} \nn\\ &+& 2 
{\hat a}_0 
{\hat a}_2 \frac{g^4_{SM}}{16 \pi^2}+ 2 ({\hat a}_1 {\hat a}_2 
+ {\hat a}_0 {\hat a}_3 )\C^{(6)}\frac{v^2}{\Lambda^2}  \frac{g^3_{SM}}{16 \pi^2}\nn\\
&+&\frac{v^4}{\Lambda^4}g_{SM}(2({\hat a}_0 {\hat a}_4+{\hat a}_1 {\hat a}_3) (\C^{(6)})^2\frac{g_{SM}}{16 \pi^2}+ ({\hat a}_0 {\hat a}_4)\C^{(8)})+\dots
\label{eq:expAsq}
\eea
As explained at the beginning of this section, in order to obtain a result that is finite in the UV in the dimension-$6$ SMEFT, 
we need to drop terms that are of order $\frac{v^4}{\Lambda^4}$ in the amplitude $\mathcal{A}$. However Eqs.~(\ref{eq:expA}) and (\ref{eq:expAsq}) show that at the squared amplitude level the requirement is slightly different: while dropping the terms $\sim (\C^{(6)})^2\frac{v^4}{\Lambda^4}$ would be inconsistent since they are in principle larger than the SM one-loop contributions, we have to drop the terms $\sim (\C^{(6)})^2\frac{v^4}{\Lambda^4}\frac{g^2_{SM}}{16 \pi^2}$ that are of the same order of the contributions from dimension 8 operators. From a practical point of view, however, those terms are for the most part numerically irrelevant, and simply squaring the amplitude (\ref{eq:expA}) after dropping the terms of order $\frac{v^4}{\Lambda^4}$ is a valid option. 
Lastly, we notice that the condition $\C^{(6)}\frac{v^2}{\Lambda^2}>\frac{g^3_{SM}}{16\pi^2}$ ensures that the fourth term in Eq.~(\ref{eq:expA}) 
is larger than a generic  SM 2-loop contribution.

\section{Results}
\label{sec:hh}
\subsection{$\h\rightarrow ZZ$}
In terms of the physical input parameters, $\mw,~\mz$ and $\gf$, the tree level SMEFT amplitude for 
the on-shell decay $\h\rightarrow Z^\mu(p_1)Z^\nu(p_2)$ is,
\begin{eqnarray}
{\cal{A}}^{0l,\mu\nu}&=&{\cal A}_{0,SM}\biggl\{
\biggl[T^{0l}_{SM}+{1\over \Lambda^2} 
\Sigma_i  \C_i
T_i^{0l} 
\biggr]\biggl(g^{\mu\nu}-{p_2^\mu p_1^\nu\over p_1\cdot p_2}\biggr)
\nonumber \\
&&+ \biggl[ T^{0l}_{SM}
+{1\over \Lambda^2} 
\Sigma_i  \C_i {\hat {T}}_i^{0l}\biggr] 
\biggl({p_2^\mu p_1^\nu\over p_1\cdot p_2}\biggr)\biggr\}
\, ,
\end{eqnarray}
where ${\cal {A}}_{0,SM}=2^{5/4}\sqrt{\gf}\mz^2$, $T_{SM}^{0l}=1$,  the sum is over
all of the contributing Warsaw basis coefficients $\C_i$, and the tree level SMEFT contribution is,
\begin{eqnarray}
\Sigma_iT_i^{0l}\C_i&=& {1\over\sqrt{2}\gf}\biggl({c_k\over 2}\biggr)
-{2p_1\cdot p_2\over \sqrt{2}\gf\mz^2}c_{ZZ}\, ,
\nonumber \\
\Sigma_i {\hat {T}}_i^{0l}\C_i&=&{2 \over \sqrt{2}\gf\mz^2} c_{ZZ} \, .
\end{eqnarray}
Note that the tree level amplitude depends on only $2$ combinations of coefficients,
\begin{eqnarray}
c_k&\equiv &
 {\cphid\over 2}+2\cphik+\cll-2 \chl3\, ,\nonumber \\
 c_{ZZ}&\equiv& \biggl[\cphiw
 {\mw^2\over \mz^2}+
 (1-{\mw^2\over \mz^2} )\cphib+{\mw\over \mz}
 \sqrt{1-{\mw^2\over \mz^2}} \cphiwb\biggr]\, .
\label{eq:czdef}
 \end{eqnarray} 
The combination $c_{ZZ}$ is limited from global fits to the  SMEFT contributions
to Higgs decays.  The $95\%$ confidence level limit is \cite{Falkowski:2015fla}\footnote{
Note the differing normalization of $c_{ZZ}$ from Ref. \cite{Falkowski:2015fla}.}
\begin{equation}
-1.2  \biggl({1~\tev\over \Lambda^2}\biggr) < c_{ZZ} < 1.6 \biggl({1~\tev\over \Lambda^2}\biggr) \, .
\end{equation}

The tree level decay width in the SMEFT is,
\begin{eqnarray}
\Gamma(\h\rightarrow ZZ)^{0l}_{EFT}&=&
{\beta\gf \mhsm^3\over 16\pi\sqrt{2}}\left(12
 \,x^2-4\,x+1\right)
\biggl\{1+
{1\over\sqrt{2}\gf\Lambda^2}c_k
\biggr\}\nonumber \\ &&
 +{3\,\beta\mhsm\over 4\pi}{\mz^2\over \Lambda^2} 
c_{ZZ}(2x-1)\nonumber \\ 
&=&\Gamma(\h\rightarrow ZZ)_{SM}\biggl\{1+{1\over \sqrt{2}\gf\Lambda^2}\biggl[
c_k+{24x(2x-1)\over 12 x^2-4x+1}c_{ZZ}\biggr]\biggr\} \, ,
\label{eq:lohzz}
 \end{eqnarray}
where $\beta=\sqrt{1-{4\mz^2\over \mhsm^2}}$ and $x\equiv \mz^2/\mhsm^2$.  
For $\mhsm$ between $2\mz$ and $200~\gev$ the dependence on $\mhsm$ is minimal~\cite{Giudice:2007fh}, 
 \begin{eqnarray}
 R^{0l} & \equiv & {\Gamma(\h\rightarrow ZZ)^{0l}_{EFT}\over \Gamma(\h\rightarrow ZZ)_{SM}^{0l}} 
 \nonumber \\
 &\sim &
 1+ {1\over \sqrt{2}\gf\Lambda^2}\biggl[
c_k-4 c_{ZZ}\biggr]+{\cal{O}}\biggl(\C{^2 v^4\over\Lambda^4}\biggr)\, .
 \label{eq:ron} 
 \end{eqnarray}
 Using the results  of Appendix \ref{sec:appa},
we can compare the on-shell tree level result  of Eq. (\ref{eq:ron}) with the off-shell result appropriate for the physical Higgs 
mass~\cite{Contino:2013kra,Grinstein:2013vsa,Isidori:2013cla},
\begin{eqnarray}
 R^{off}&\equiv& {\Gamma(\h\rightarrow Zf {\overline f})^{0l}\over 
 \Gamma(\h\rightarrow Z f {\overline f})_{SM} }\nonumber \\
 &\sim & 1+ {1\over \sqrt{2}\gf\Lambda^2}\biggl[
c_k-.97 c_{ZZ}\biggr]+{\cal{O}}\biggl(\C{^2 v^4\over\Lambda^4}\biggr)\, .
 \label{eq:roff}
 \end{eqnarray}
 Comparing Eqs. (\ref{eq:ron}) and  (\ref{eq:roff}), we note that  the off-shell effects are large~\cite{Boselli:2017pef}. 

In the SMEFT, the dimension-$6$ NLO results contain both 
 the complete SM $1-$loop electroweak corrections to $\h\rightarrow ZZ$  and the one-loop
corrections from the dimension-$6$ SMEFT operators.  
The complete, renormalized NLO amplitude is,
\begin{eqnarray}
{\cal A}^{\mu\nu}_{NLO}={\cal A}^{0l,\mu\nu}+{\cal A}^{1l,\mu\nu}\, .
\end{eqnarray}
The one-loop SMEFT contribution to the  amplitude can be expanded as, 
\bea
{\cal{A}}^{1l,\mu\nu}&=&{\cal A}_{0,SM}\biggl\{\biggl(T^{1l}_{SM}+\mathcal{F}_g \log\biggl({\Lambda^2\over \mz^2}\biggr)+{(1~TeV)^2\over \Lambda^2}\sum_i
T_i^{1l}\C_i \biggr) \biggl(g^{\mu\nu}-\frac{ p_1^\nu p_2^\mu}{p_1\cdot p_2}\biggr)\nn\\&+&\biggl({\hat {T}}^{1l}_{SM}+\mathcal{F}_p \log\biggl({\Lambda^2\over \mz^2}\biggr)+{(1~TeV)^2\over\Lambda^2}
\sum_i {\hat {T}}_i^{1l}\C_i\biggr) \frac{ p_1^\nu p_2^\mu}{p_1\cdot p_2}\biggr\}\, ,
\label{eq:ampres}
\eea
where the terms $\mathcal{F}_g \log\biggl({\Lambda^2\over \mz^2}\biggr)$ and $\mathcal{F}_p \log\biggl({\Lambda^2\over \mz^2}\biggr)$ contain the residual dependence on the renormalization scale, due to our choice of renormalization scheme. 
  Retaining terms to ${\cal O}\biggl({v^2\over \Lambda^2}\biggr)$, we
   parameterize the exact  one-loop SMEFT
result,\footnote{This fit is valid for $\mhsm < 2 M_t$.}
\begin{eqnarray}
T_i^{1l}&=& a_{0,i}+a_{1,i}\hoz+a_{2,i}\biggl(\hoz\biggr)^2+a_{3,i}\log\biggl(\hoz\biggr)+
a_{4,i}\log\biggl(\thz\biggr)\nonumber \\
{\hat T} _i^{1l}&=& b_{0,i}+b_{1,i}\hoz+b_{2,i}\biggl(\hoz\biggr)^2+b_{3,i}\log\biggl(\hoz\biggr)+
a_{4,i}\log\biggl(\thz\biggr)\, . 
\label{eq:tabdef}
\end{eqnarray}
Numerical values for the fit parameters, together with the analytical expressions for $\mathcal{F}_g$ and $\mathcal{F}_p$ are given in Appendix \ref{sec:appfit}. 

The SM result is easily recovered,
\begin{equation}
{\cal A}^{1l,\mu\nu}_{SM}=
{\cal A}_{0,SM}
\biggl\{ T^{1l}_{SM} \biggl(g^{\mu\nu}-\frac{ p_1^\nu p_2^\mu}{p_1\cdot p_2}\biggr)
+{\hat {T}}^{1l}_{SM}
 \frac{ p_1^\nu p_2^\mu}{p_1\cdot p_2}\biggr\}\, ,
 \end{equation}
where $T_{SM}^{1l}=a_{0,SM}$ and ${\hat T}^{1l}_{SM}=b_{0,SM}$. We have verified
that our one-loop electroweak SM corrections are in agreement with previous results~\cite{Kniehl:1990mq}. 
\begin{figure}
  \centering
\includegraphics[width=0.45\textwidth]{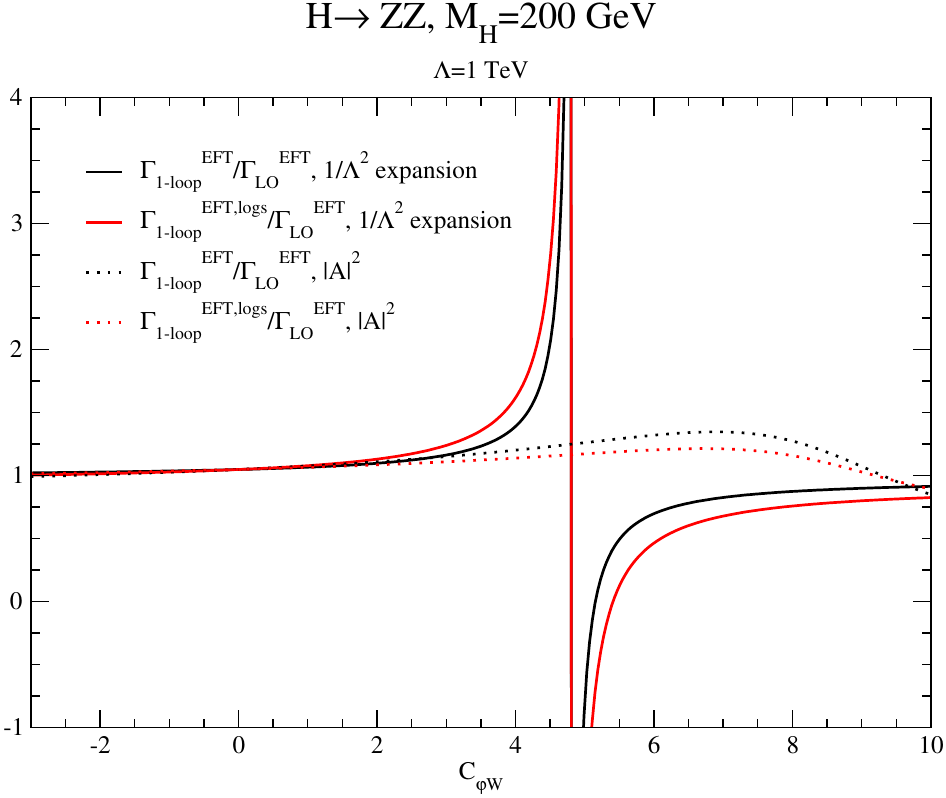}
 \caption{Dependence of $H\rightarrow ZZ$ decay width on $\cphiw$ in different expansions
 as explained in the text.  \label{fg:gam2}}
\end{figure}

\begin{figure}
  \centering
\includegraphics[width=0.45\textwidth]{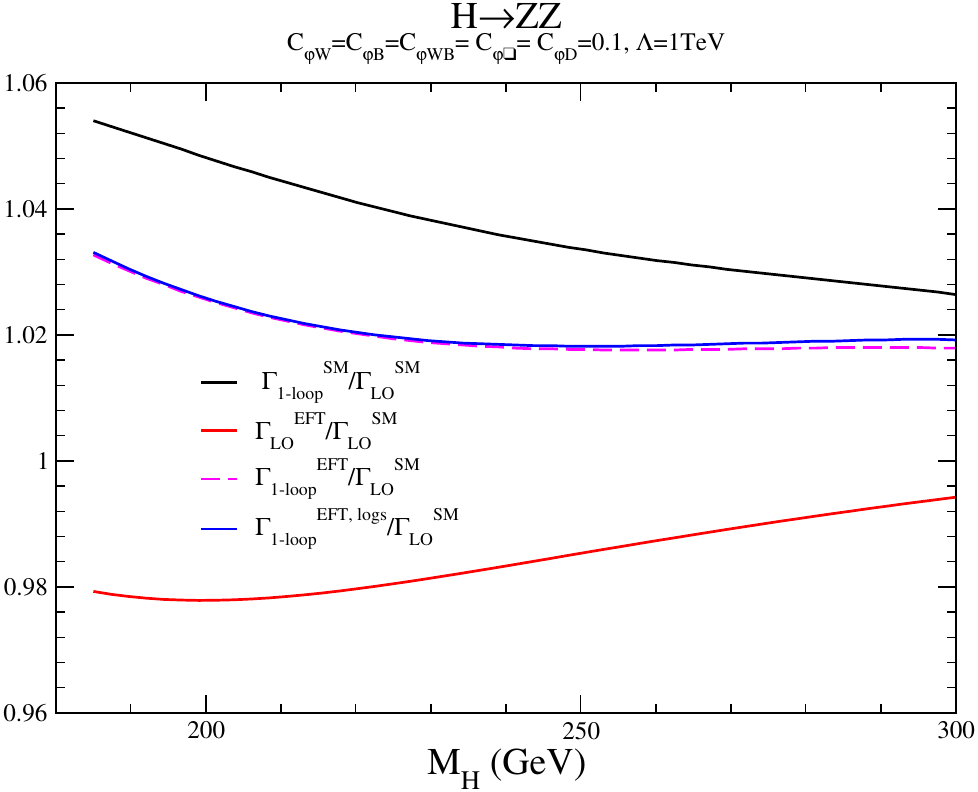}
\includegraphics[width=0.45\textwidth]{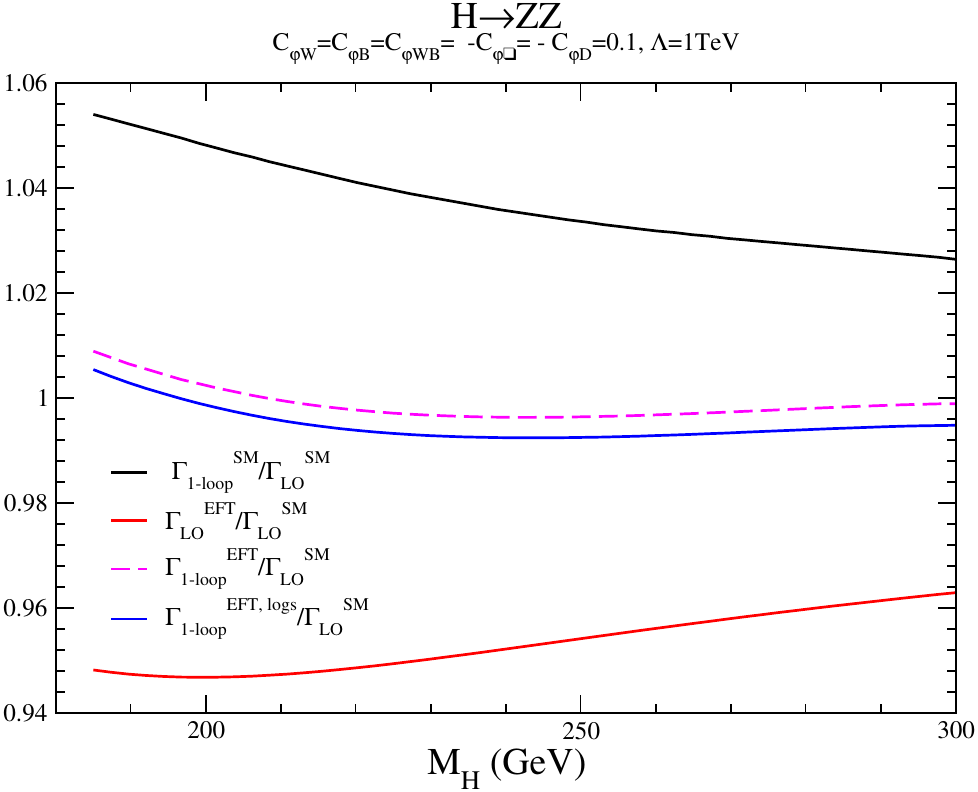}
 \caption{Dependence of the $H\rightarrow ZZ$ decay width on SMEFT coefficients that
 contribute at tree level.  Note that the curve labelled $1-$loop EFT is the complete SMEFT result
 and contains both the one-loop SM result and the one-loop contribution from the dimension-6 SMEFT coefficients.
 Coefficients not specified are set to $0$.}
  \label{fg:gam3}
\end{figure}

In Fig. \ref{fg:gam2}, we illustrate the dependence of the decay widths on the terms retained in the 
expansion of Eq. (\ref{eq:expAsq}).
The curves labelled  ``$1/\Lambda^2$ expansion" drop the ${\hat a}_1^2$ and $({\hat a}_0{\hat a}_4+
{\hat a}_1{\hat a}_3)$
contributions in Eq. (\ref{eq:expAsq}). It is apparent that for large values of the $\C$ (here $\cphiw$), the expansion is
nonsense.   For $\C_{\phi W}> 2$, the dimension-$6$ approximation breaks  down and the dimension-$6$
approximation to the amplitude-squared becomes negative. The curves labelled $\mid {\cal A}\mid^2$ contain all of the terms in the square of  Eq. (\ref{eq:expA})
 (with ${\hat a}_4\rightarrow
0$) and are  well behaved for all $\C$. For $\cphiw < 2$, the full amplitude-squared is 
well approximated by the terms linear in the coefficient functions and the SMEFT d=6 approximation is valid.
We have checked that this condition is satisfied for the following plots in this section.

\begin{figure}
  \centering
\includegraphics[width=0.45\textwidth]{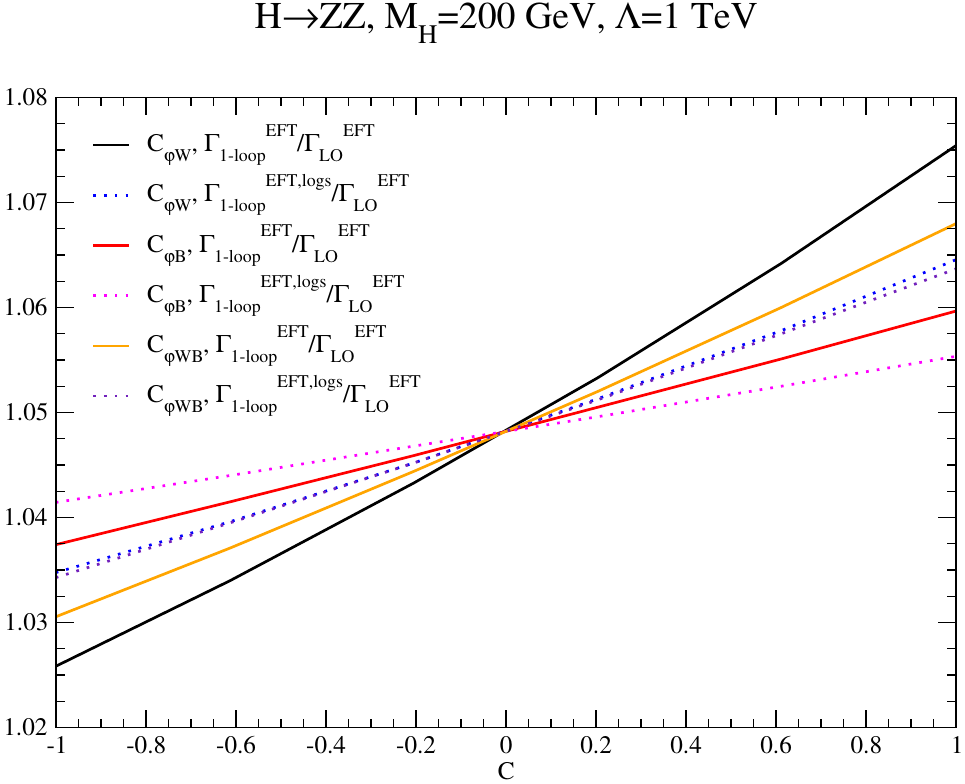}
\includegraphics[width=0.45\textwidth]{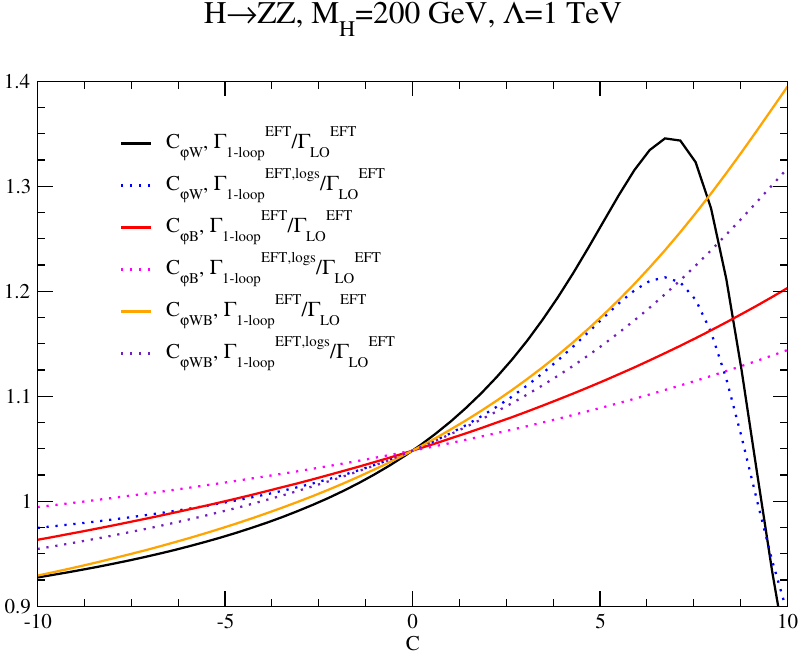}
\caption{Dependence of $H\rightarrow ZZ$ decay width on SMEFT coefficients.
Coefficients are varied one at a time and 
coefficients not specified are set to $0$.}
  \label{fg:cuw}
\end{figure}

\begin{figure}
  \centering
\includegraphics[width=0.45\textwidth]{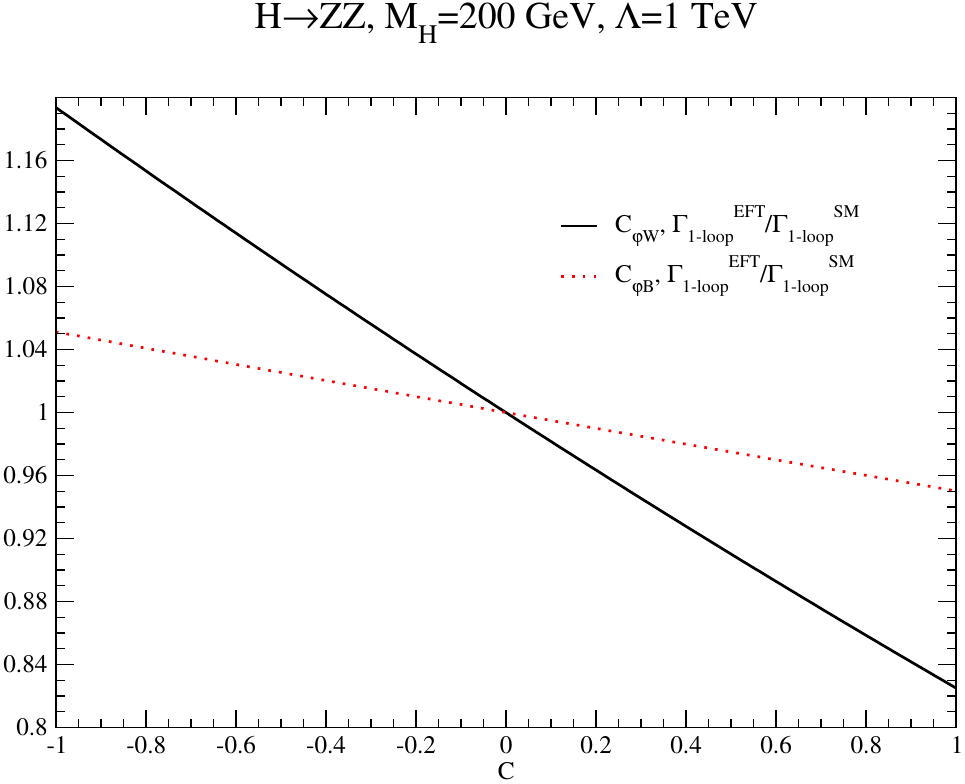}
\caption{Same as Fig. \ref{fg:cuw}, but normalized to $\Gamma_{1-loop}^{SM}$.
  \label{fg:sm}}
\end{figure}

In Fig. \ref{fg:gam3}, we show the contribution relative to the LO SM prediction for representative values of the SMEFT 
coefficients that contribute at tree level.  Choosing all coefficients positive, for the parameters we have chosen,
inclusion of the tree level SMEFT coefficients decreases the rate by $\sim2\%$ (red curve) .
The SM one-loop corrections (black curve) 
increase the rate by roughly $5\%$,  leading to a partial cancellation between the 1-loop SM and tree level
SMEFT contributions.    This makes it clear that global fits to SMEFT coefficients that do not contain
the electroweak corrections cannot be more accurate than $\sim {\cal{O}}(5\%)$. The one loop corrections  from the
the SMEFT operators are much  smaller than the SM electroweak
corrections.   In this example, the complete NLO SMEFT calculation is well approximated by
including only the logarithmic contributions from the SMEFT coefficients. 
On the RHS of  Fig. \ref{fg:gam3}, we flip the sign of the SMEFT coefficients, which increases their 
numerical significance.  Note that the dependence on $\mhsm$ is rather mild. 

\begin{figure}
  \centering
\includegraphics[width=0.45\textwidth]{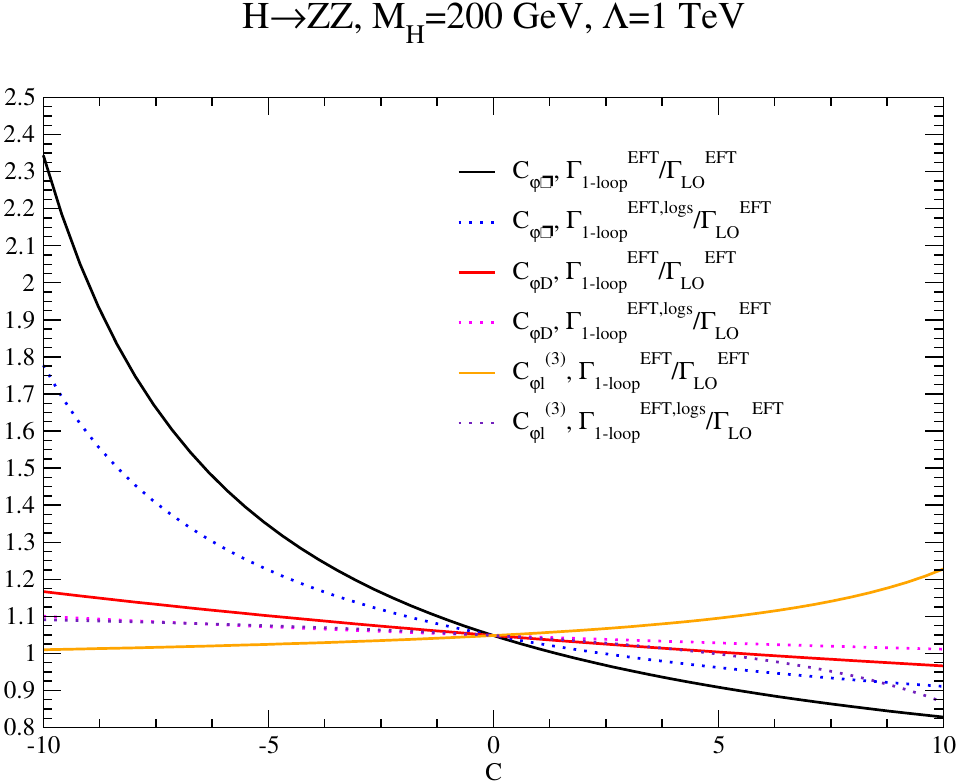}
\includegraphics[width=0.45\textwidth]{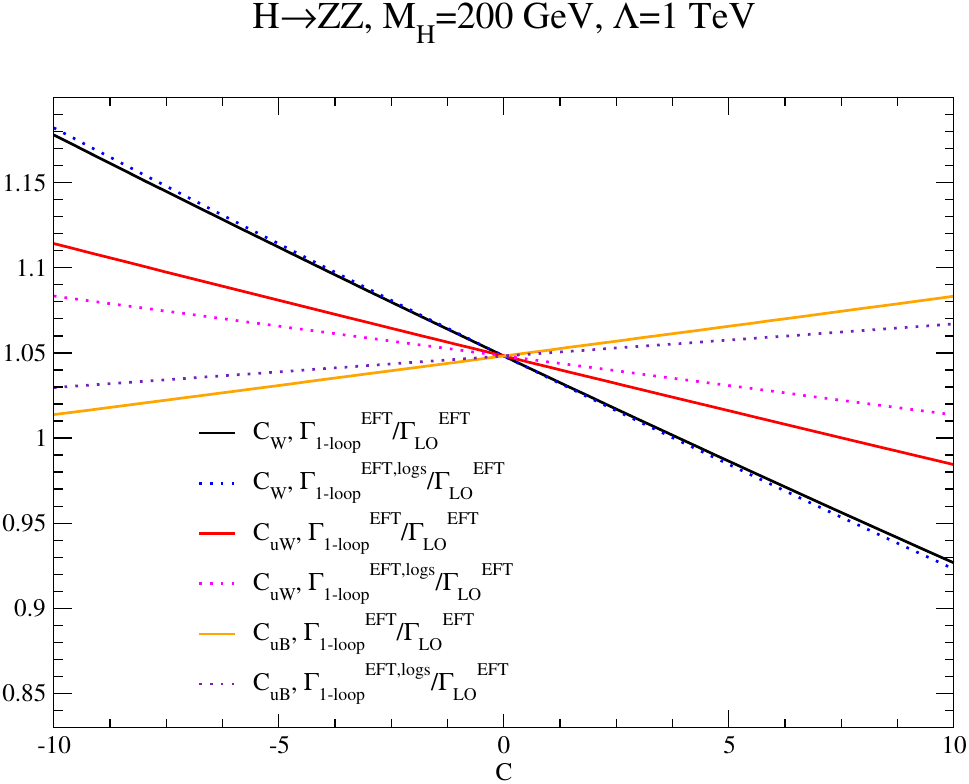}
 \caption{Dependence of the one-loop corrected SMEFT width on  various $\C$ for 
$\mhsm=200~GeV$. 
  \label{fg:cll}}
\end{figure}

In Figs.   \ref{fg:cuw} and   \ref{fg:cll} , we show the dependence on SMEFT coefficients as a function of the $\C$ for fixed $\mhsm=200~\gev$,
normalized to the LO SMEFT result.    
 Fig. \ref{fg:cuw} shows the dependence on coefficients that enter at tree level and in 
Fig. \ref{fg:sm}, we replot the same results, normalized to the one-loop SM.    
The change from the one-loop SM results in the SMEFT is of ${\cal {O}}(5\%)$ 
for $\cphib\sim {\cal{O}}(-1)$ and  for  $\cphiw \sim {\cal{O}}(-1)$ this change is ${\cal {O}}(20\%)$.  These
values are consistent with current fits to LHC Higgs decays to $ZZ$. 

Fig. \ref{fg:cll} shows the dependence on a selection of coefficient functions that do not enter at tree level.  An interesting feature
of our results is that they can be used to obtain limits on coefficients that first enter at loop level.  For example,
from a global fit~\cite{Berthier:2016tkq,Falkowski:2015fla,Brivio:2017bnu},
\begin{eqnarray}
\C_W&=& {\Lambda^2\over v^2}(1.14\pm .68)\nn\\
&\rightarrow& (18.8\pm 11.2)\biggl ({\Lambda^2\over 1~\tev^2}\biggr)\, .
\end{eqnarray}
Such a large value of $\C_W$ would increase the LO SMEFT width to $ZZ$ by $\sim 20\%$ as observed in Fig.
\ref{fg:cll}.  Loop corrections to Higgs decays therefore have the possibility of new constraints on the SMEFT
coefficients. 

Our corrections to the on-shell $\h\rightarrow ZZ$  one-loop SMEFT result must be considered as a first step in
a full SMEFT calculation for $\h\rightarrow Zf {\overline f}$.  Our results suggest, however, that the 
results of Figs.   \ref{fg:cuw} and \ref{fg:cll}  can be thought of as $K$ factors that multiply the LO SMEFT
off-shell result of Eq. (\ref{eq:roff}).  The numerical size of our results implies that a full SMEFT calculation
is needed for $\h\rightarrow Zf {\overline f}$ in order to obtain reliable results.  The SM electroweak corrections
for $\h\rightarrow Zf {\overline f}$ 
are known and can be implemented using the
PROPHECY4F program~\cite{Bredenstein:2006ha,Bredenstein:2006nk} .

\subsection{$\h\rightarrow Z\gamma$}

The one-loop SMEFT results for $\h\rightarrow Z^\mu(p_1) \gamma^\nu(p_2)$ 
can be obtained in a straightforward manner from the results of the previous section.  At tree
level, there is  an SMEFT contribution,  while the SM contribution begins at one- loop.  The Ward identity for the photon requires 
$ p^\nu\cdot {\cal{A}}_{\mu\nu}=0$, so the
tensor structure is fixed, 
\begin{equation}
{\cal{A}}^{\mu\nu}_{Z\gamma}=\biggl[
{\cal{A}}_{Z\gamma, EFT}^{0l}+{\cal{A}}_{Z\gamma,SM }^{1l}+
{\cal{A}}_{Z\gamma,EFT}^{1l}\biggr]
\biggl(g_{\mu\nu}- {p_1^\nu p_2^\mu\over p_1\cdot p_2} \biggr)\, .
\end{equation}
The tree level SMEFT contribution is 
\begin{equation}
{\cal {A}}_{Z\gamma,EFT} ^{0l} =  - 2^{3/4}\gf^{-1/2}
\biggl({\mhsm^2-\mz^2\over\Lambda^2}\biggr) c_{Z\gamma} \, ,
\end{equation}
with 
\begin{equation}
c_{Z\gamma}\equiv {\mw\over \mz}\sqrt{1-{\mw^2\over \mz^2}}\biggl(\cphiw-\cphib\biggr)+\frac12\biggl(1-2{\mw^2\over
\mz^2}\biggr)\cphiwb\, .
\label{eq:czgdef}
\end{equation}
Numerically, for $\mhsm=125~\gev$ and $\Lambda=1~\tev$,
\begin{equation}
{\cal {A}}_{Z\gamma,EFT} ^{0l} =-1.5~\gev\biggl(\cphiw-\cphib-0.67\cphiwb\biggr)\biggl({1~\tev\over\Lambda^2}\biggr)\, .
\label{eq:gztree}
\end{equation}

The SM contribution is well-known,
\begin{eqnarray}
{\cal{A}}_{Z\gamma,{SM}}^{1l}&=&
\biggl(
{\mhsm^2-\mz^2\over 2}\biggr) {\mw^2\over \pi^2}\gf^{3/2} 2^{-1/4} \sqrt{1-{\mw^2\over \mz^2}}
\biggl\{\Sigma _f N_c {Q_f \mz\over \mw} v_f A_{1/2}(\tau_f,\lambda_f)\nonumber \\
&&
+A_1(\tau_W,\lambda_W)\biggr\} 
\nonumber \\
&=& 0.209~\gev\,{\hbox{for}}~\mhsm=125~\gev\, ,
\label{eq:zgsm}
\end{eqnarray}
where the sum is over all fermions, $N_c=1(3)$ for leptons (quarks), $\tau_i=4M_i^2/\mhsm^2$, $\lambda_i=4M_i^2/\mz^2$, 
$v_f=2 T^3_f-4 Q_f \biggl(1-{\mw^2\over\mz^2}\biggr)$,
and analytic expressions for the functions $A_1$ and $A_{1/2}$ can be found in Refs.~ \cite{Cahn:1978nz,Bergstrom:1985hp,Gunion:1989we}

We report our one-loop  SMEFT corrections\footnote{In the previous version of this paper we calculated the corrections to the vacuum polarization using the light quark masses instead of the more accurate $\Delta\alpha_{had}^{(5)}$. The difference between the two calculations is of order percent and noticeable only in the corrections proportional to $C_{\phi W}$, $C_{\phi B}$ and $C_{\phi WB}$ since they contribute already at LO. We thank the authors of Ref. ~\cite{Dedes:2019bew} for 
pointing out the numerical effect of $\Delta\alpha_{\rm had}^{(5)}$ and for noticing a typo in Eqs.~(\ref{eq:czgdef}) and (\ref{eq:gztree}). } to the $H\rightarrow Z\gamma$ amplitude, $\mathcal{A}^{1l}_{Z\gamma}$ ,
numerically for $\mhsm=125~\gev$,

\bea
{\cal {A}}_{Z\gamma,EFT}^{1l}&=&
\frac{(1 TeV)^2}{\Lambda^2} \biggl\{-0.038\,\C_{\phi l}^{(3)}+0.00185(\C_{\phi q}^{(1)}-\C_{\phi q}^{(3)}+\C_{\phi u})-0.0126\,\C_{\phi D}\nn\\ &&
+0.0127\,\C_{\phi\square}+0.000753\,\C_{u \phi}+0.019\,\C_{ll}+(0.0564-0.0362 \log \frac{\Lambda^2}{M_Z^2})\C_{\phi B}
\nn\\&&
+(-0.0502+0.0154 \log \frac{\Lambda^2}{M_Z^2})\C_{\phi W}+(-0.0201-0.0269 \log \frac{\Lambda^2}{M_Z^2}) \C_{\phi WB}
\nn\\&&
+(-0.00999+0.0042 \log \frac{\Lambda^2}{M_Z^2}) \C_{uB}+(0.0669-0.0295 \log \frac{\Lambda^2}{M_Z^2}) \C_{uW}
\nn\\&& 
+(0.00559-0.0213 \log \frac{\Lambda^2}{M_Z^2}) \C_{W}\biggr\}~\gev\, .
\label{eq:zg1loop}
\eea
An interesting feature of Eq. (\ref{eq:zg1loop})
is the dependence on coefficients not arising at tree level.

\begin{figure}
  \centering
\includegraphics[width=0.45\textwidth]{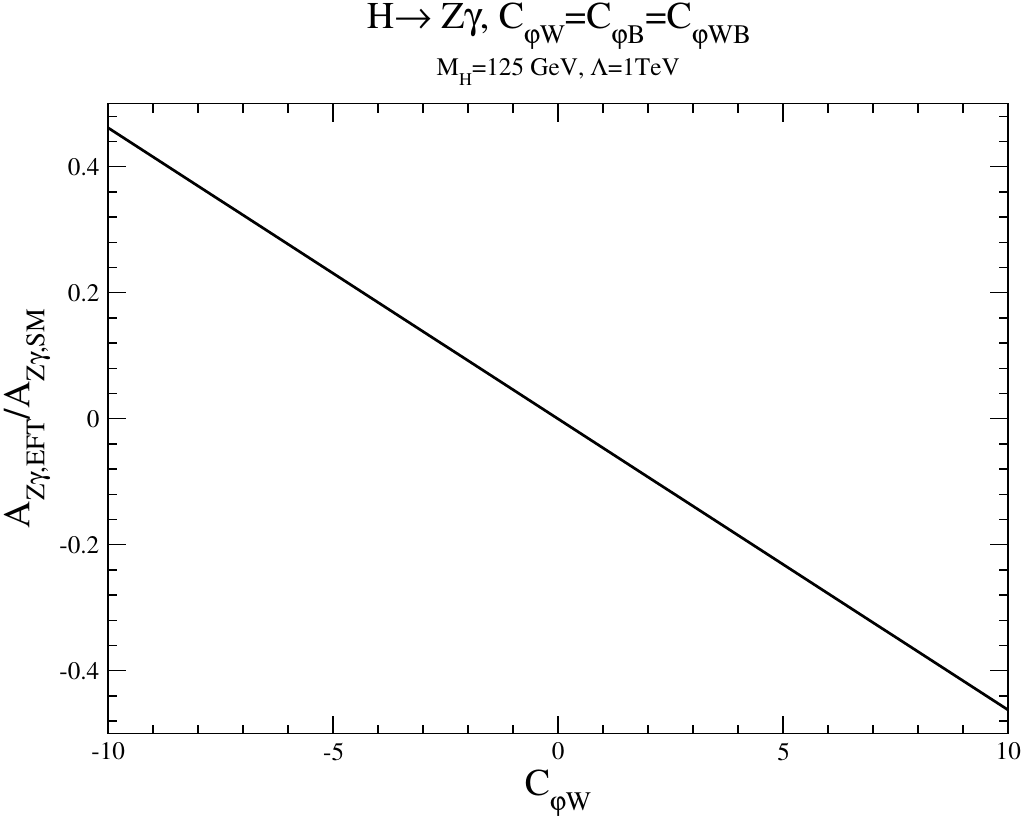}
\caption{
Contribution to $H\rightarrow Z\gamma$ in the SMEFT.  
The vertical axis is the ratio of the one-loop contribution of
Eq. (\ref{eq:zg1loop}) to the SM one-loop result of Eq.~ (\ref{eq:zgsm}).
  \label{fg:hzg}
  }
\end{figure}

\label{sec:hgg}
The contribution from the SMEFT operators to $H\rightarrow Z\gamma$ at $1-$ loop is much smaller than the SM $1-$loop contribution,
due to the ${v^2\over \Lambda^2}$ suppression.    The contribution to the amplitude, Eq.~(\ref{eq:zg1loop}) is shown in Fig. \ref{fg:hzg}

\section{Conclusions}
We have computed the one-loop electroweak corrections in the SMEFT to the on-shell $\h\rightarrow ZZ$ process.  Numerically, the logarithmic
SMEFT contributions dominate over the finite NLO contributions for most of  parameter space. 
Appendix \ref{sec:appfit} contains a numerical fit to the finite NLO SMEFT contributions and Appendix \ref{sec:appf} has analytic 
results for the logarithmic contributions to $\h\rightarrow ZZ$.  Our complete result can be obtained from the ancillary files posted 
with this paper and at
\url{https://quark.phy.bnl.gov/Digital_Data_Archive/dawson/zzNLO_18}.
The calculation of the on-shell decay is a first step towards a full NLO
calculation of the physical $\h\rightarrow Z f {\overline f}$ process and our results can be used to approximate a $K$ factor for the
SMEFT $\h\rightarrow Z f {\overline f}$ decay.  

The full $\h\rightarrow Z\gamma$ SMEFT NLO result is presented as a by-product of our calculation.  Finally, the
complete result for the one-loop SMEFT renormalization of $G_\mu$ is given for the first time. 

\section*{Acknowledgements}
We thank Giuseppe Degrassi for discussions. After the publication of this paper,  Ref.~\cite{Dedes:2019bew} came out. We thank its authors, Athanasios Dedes, Kristaq Suxho and Lampros Trifyllis, for noticing a typo in  Eqs. (38) and (39), and for pointing out the impact of $\Delta\alpha_{\rm had}^{(5)}$ in the calculation of the NLO corrections to $\C_{\phi W}, \C_{\phi B}$ and $\C_{\phi W B}$.
S.D.   and P.P.G are supported by the U.S. Department of Energy under Grant Contract  de-sc0012704.  

\appendix
\section{Off-shell Production}
\label{sec:appa}
The decay width for  the off-shell decay, $H\rightarrow f_1(p_1) f_2(p_2) Z(p_3)$, is
\begin{eqnarray}
\Gamma&=&\int_0^{(M_H-M_Z)^2} dq^2 \int dm_{23}^2\,{\mid A\mid^2\over 256\pi^3 M_H^3}\, ,
\end{eqnarray}
where $m_{ij}=(p_i+p_j)^2$, $m_{12}^2\equiv q^2$,and $m_{12}^2+m_{23}^2+m_{13}^2=M_H^2+M_Z^2$,
$\lambda(M_H^2,M_Z^2,q^2)\equiv q^4-2q^2(M_H^2+M_Z^2)+(M_H^2-M_Z^2)^2$,
$m_{23}^2\mid_{max,min}\equiv {1\over 2}
\biggl(M_H^2+M_Z^2-q^2\pm \sqrt{\lambda}\biggr)$.
We write the amplitude-squared  to ${\cal O}({1\over \Lambda^2})$ in the SMEFT as,
\begin{eqnarray}
\mid A_{EFT}\mid^2&=& \mid A_{SM}\mid^2+\mid \delta A_{EFT}\mid^2
+{\cal{O}}\biggl(\C^2{v^4\over \Lambda^4}\biggr)
\end{eqnarray}
where,
\begin{eqnarray}
\mid A_{SM}\mid^2&=&
32\,(g_L^{\, 2}+g_R^{2})\,G_F^2\,
M_Z^4\biggl [
{2 M_Z^2q^2 -m_{13}^2 q^2 
-M_H^2 M_Z^2+m_{13}^2M_Z^2
 +m_{13}^2\,M_H^2- m_{13}^4\over
 (q^2-M_Z^2)^2+\Gamma_Z^2M_Z^2}\biggr]
 \nonumber \\
\mid \delta A_{EFT}\mid^2&=&\mid A_{SM}\mid^2
{1\over \sqrt{2}G_F\Lambda^2}c_k
 \nonumber \\ &&
+64 (g_L^2+g_R^2) \frac{\sqrt{2} G_F}{\Lambda^2} M_Z^4
{q^2(q^2+M_Z^2-M_H^2)
\over (q^2-M_Z^2)^2+\Gamma_Z^2 M_Z^2}\, c_{ZZ} \, .
  \label{eq:treeamp}
\end{eqnarray}
Note that we have not included effects due to possible anomalous
$\h \rightarrow Z\gamma$ vertices here, although they are included in the results of 
Sec. \ref{sec:hgg}.

Integrating over $dm_{23}^2$, 
\begin{eqnarray}
{d\Gamma\over dq^2}\mid_{SM}&=&
(g_L^{\, 2}+g_R^{2})\,G_F^2\,\sqrt{\lambda(M_H^2,M_Z^2,q^2)}{M_Z^4\over 48 \pi^3 M_H^3}
\biggl[{(12 M_Z^2 q^2+\lambda(M_H^2,M_Z^2,q^2))
  \over
 (q^2-M_Z^2)^2+\Gamma_Z^2M_Z^2}
\biggr]
\nonumber \\
{d\Gamma\over dq^2}\mid_{EFT}&=&
 {d\Gamma\over dq^2}\mid_{SM}
\biggl\{
1+
{1\over \sqrt{2}G_F\Lambda^2} c_k\biggr\}
 \nonumber \\ &&
+
(g_L^{\, 2}+g_R^{2})\,\frac{G_F}{\Lambda^2}\,\sqrt{\lambda(M_H^2,M_Z^2,q^2)}{M_Z^4\over 2\sqrt{2} \pi^3 M_H^3}
\nonumber \\ &&\cdot 
{q^2(3M_Z^2+M_H^2)-(M_Z^2-M_H^2)^2+\lambda(M_H^2,M_Z^2,q^2)
\over (q^2-M_Z^2)^2+\Gamma_Z^2 M_Z^2}
c_{ZZ}\, .
\end{eqnarray}

\begin{figure}
  \centering
\includegraphics[width=0.45\textwidth]{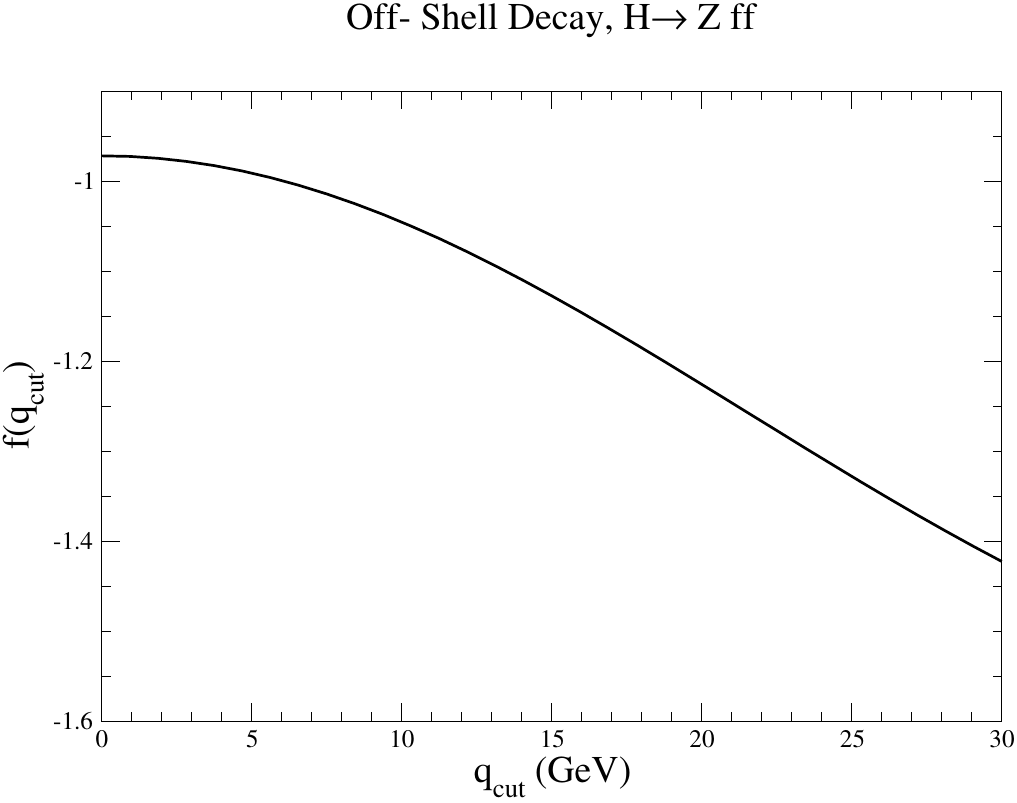}
 \caption{Dependence of the off-shell decay rate on the $q_{cut}$ defined in Eq. \ref{eq:resof}.}
  \label{fg:q2}
\end{figure}

Finally, integrating over $q^2$ and setting $\mhsm=125~\gev$, 
\begin{eqnarray}
R^{off}
& \equiv &  {\int~dq^2~d\Gamma/dq^2\mid_{EFT}\over \int~dq^2~d\Gamma/dq^2_{SM}}
\nonumber \\
&\sim &1+{1\over\sqrt{2} G_F\Lambda^2}\biggl\{
c_k- .97c_{ZZ} \biggr\}
 \label{eq:reso}
 \end{eqnarray}
 This is in agreement with 
 Refs. \cite{Brivio:2017vri,Grinstein:2013vsa,Buchalla:2013mpa}.  Note that if we require a minimum $q^2\equiv q^2_{cut}$ , the result is altered. Define,
  \begin{eqnarray}
R^{off}(q^2_{cut})
& \equiv &  {\int_{q^2_{cut}}^{(\mhsm-\mz)^2}~dq^2~d\Gamma/dq^2\mid_{EFT}
\over \int_{q^2_{cut}}^{(\mhsm-\mz)^2}~dq^2~d\Gamma/dq^2_{SM}}
\nonumber \\
&=&1+{1\over\sqrt{2} G_F\Lambda^2}\biggl\{
c_k+f(q_2^{cut}) c_{ZZ} \biggr\}  \, .
 \label{eq:resof} 
 \end{eqnarray}
 The effects of the $q^2_{cut}$ are shown in Fig. \ref{fg:q2}.
 Refs. \cite{Contino:2013kra,Boselli:2017pef} note the the numerically large effect of off-shell $Z$'s, which is seen clearly in
 the sensitivity to the $q^2_{cut}$. 

\section{Numerical values for the on-shell decay $\h\rightarrow ZZ$}
\label{sec:appfit}

Numerical values for the $1-$loop  SMEFT on-shell decay $\h\rightarrow ZZ$ in terms of the parameterization of Eq. \ref{eq:tabdef}
are given in Tables 1-3. 
\bea
\left(
\begin{array}{cccccc}
 & a_0 & a_1 & a_2 & a_3 & a_4 \\
 \text{SM} & -3.09\times 10^{-2} & -1.41\times 10^{-2} & 5.43\times 10^{-4} & 4.93\times 10^{-2} & 1.57\times 10^{-2} \\
 \C_W & -3.08\times 10^{-3} & -8.96\times 10^{-5} & 4.76\times 10^{-5} & 1.43\times 10^{-4} & 8.29\times 10^{-4} \\
 \C_\phi & -2.12\times 10^{-3} & -2.11\times 10^{-4} & 6.59\times 10^{-6} & 1.41\times 10^{-3} & 1.2\times 10^{-4} \\
 \C_{\phi\square} & 2.99\times 10^{-3} & 2.11\times 10^{-4} & 2.9\times 10^{-5} & -1.14\times 10^{-4} & 1.23\times 10^{-3} \\
 \C_{\phi D} & 1.42\times 10^{-3} & -1.63\times 10^{-5} & 7.09\times 10^{-6} & 4.86\times 10^{-4} & 1.51\times 10^{-4} \\
 \C_{u\phi} & 1.56\times 10^{-3} & -6.49\times 10^{-6} & 1.3\times 10^{-6} & 3.29\times 10^{-5} & -4.65\times 10^{-4} \\
 \C_{\phi W} & 1.14\times 10^{-2} & 8.69\times 10^{-4} & -1.75\times 10^{-4} & -1.02\times 10^{-2} & -7.56\times 10^{-4} \\
 \C_{\phi B} & 3.52\times 10^{-3} & 2.64\times 10^{-5} & -8.25\times 10^{-5} & -2.02\times 10^{-3} & -5.46\times 10^{-4} \\
 \C_{\phi WB} & 5.88\times 10^{-3} & -1.63\times 10^{-3} & -5.93\times 10^{-5} & 1.62\times 10^{-3} & -8.97\times 10^{-4} \\
 \C_{uW} & -1.72\times 10^{-3} & 7.28\times 10^{-4} & -4.99\times 10^{-6} & -5.86\times 10^{-5} & 1.36\times 10^{-4} \\
 \C_{uB} & 9.2\times 10^{-4} & -3.9\times 10^{-4} & 2.67\times 10^{-6} & 3.14\times 10^{-5} & -7.3\times 10^{-5} \\
 \C_{\phi l}^{(1)} & 6.82\times 10^{-7} & 0 & 0 & 0 & 0 \\
 \C_{\phi l}^{(3)} & 3.56\times 10^{-3} & 2.56\times 10^{-3} & -9.87\times 10^{-5} & -8.97\times 10^{-3} & -2.85\times 10^{-3} \\
 \C_{\phi e} & 6.26\times 10^{-5} & 0 & 0 & 0 & 0 \\
 \C_{\phi q}^{(1)} & 4.45\times 10^{-3} & -2.23\times 10^{-5} & -9.49\times 10^{-7} & -1.64\times 10^{-5} & 5.43\times 10^{-5} \\
 \C_{\phi q}^{(3)} & -6.06\times 10^{-3} & 2.23\times 10^{-5} & 9.49\times 10^{-7} & 1.64\times 10^{-5} & -5.43\times 10^{-5} \\
 \C_{\phi u} & -4.71\times 10^{-3} & 5.2\times 10^{-5} & 2.14\times 10^{-6} & 3.67\times 10^{-5} & -1.2\times 10^{-4} \\
 \C_{\phi d} & 6.26\times 10^{-5} & 0 & 0 & 0 & 0 \\
\C_{ll} & -3.17\times 10^{-3} & -1.28\times 10^{-3} & 4.93\times 10^{-5} & 4.49\times 10^{-3} & 1.43\times 10^{-3} \\
\C_{lq}^{(3)} & 8.9\times 10^{-4} & 0 & 0 & 0 & 0 \\
\end{array}
\right)
\nonumber \label{eq:fitzz1}
\eea
\noindent{Table 1: Numerical values for the coefficients defined in Eq. \ref{eq:tabdef} relevant for the on-shell process $\h\rightarrow ZZ$.}

\bea
\left(
\begin{array}{cccccc}
 & b_0 & b_1 & b_2 & b_3 & b_4 \\
 \text{SM} & -2.68\times 10^{-2} & -6.16\times 10^{-3} & 2.83\times 10^{-4} & 2.49\times 10^{-2} & 1.03\times 10^{-2} \\
 \C_W & -1.6\times 10^{-3} & -4.44\times 10^{-4} & 1.62\times 10^{-5} & 1.76\times 10^{-3} & 3.44\times 10^{-4} \\
 \C_\phi & -2.04\times 10^{-3} & -1.88\times 10^{-4} & 5.84\times 10^{-6} & 1.26\times 10^{-3} & 1.06\times 10^{-4} \\
 \C_{\phi\square} & 3.2\times 10^{-3} & 7.73\times 10^{-4} & 1.15\times 10^{-5} & -1.74\times 10^{-3} & 8.63\times 10^{-4} \\
 \C_{\phi D} & 1.51\times 10^{-3} & 1.23\times 10^{-4} & 3.13\times 10^{-6} & 8.87\times 10^{-5} & 4.19\times 10^{-5} \\
 \C_{u\phi} & 1.52\times 10^{-3} & -7.6\times 10^{-6} & 1.08\times 10^{-6} & 2.86\times 10^{-5} & -4.49\times 10^{-4} \\
 \C_{\phi W} & 6.37\times 10^{-3} & 1.07\times 10^{-3} & -3.92\times 10^{-5} & -5.3\times 10^{-3} & -8.58\times 10^{-4} \\
 \C_{\phi B} & 7.33\times 10^{-4} & 9.47\times 10^{-5} & -3.39\times 10^{-6} & -4.85\times 10^{-4} & -7.1\times 10^{-5} \\
 \C_{\phi WB} & 8.71\times 10^{-4} & -1.19\times 10^{-4} & 6.47\times 10^{-6} & 1.17\times 10^{-5} & 1.92\times 10^{-4} \\
 \C_{uW} & 1.09\times 10^{-5} & -4.3\times 10^{-8} & -1.47\times 10^{-8} & -3.18\times 10^{-7} & 1.48\times 10^{-6} \\
 \C_{uB} & -5.85\times 10^{-6} & 2.3\times 10^{-8} & 7.87\times 10^{-9} & 1.7\times 10^{-7} & -7.93\times 10^{-7} \\
 \C_{\phi l}^{(1)} & 6.82\times 10^{-7} & 0 & 0 & 0 & 0 \\
 \C_{\phi l}^{(3)} & 2.8\times 10^{-3} & 1.12\times 10^{-3} & -5.14\times 10^{-5} & -4.52\times 10^{-3} & -1.88\times 10^{-3} \\
 \C_{\phi e} & 6.26\times 10^{-5} & 0 & 0 & 0 & 0 \\
 \C_{\phi q}^{(1)} & 4.36\times 10^{-3} & -3.33\times 10^{-5} & -1.75\times 10^{-6} & -3.11\times 10^{-5} & 1.06\times 10^{-4} \\
 \C_{\phi q}^{(3)} & -5.97\times 10^{-3} & 3.33\times 10^{-5} & 1.75\times 10^{-6} & 3.11\times 10^{-5} & -1.06\times 10^{-4} \\
 \C_{\phi u} & -4.71\times 10^{-3} & 3.33\times 10^{-5} & 1.75\times 10^{-6} & 3.11\times 10^{-5} & -1.06\times 10^{-4} \\
 \C_{\phi d} & 6.26\times 10^{-5} & 0 & 0 & 0 & 0 \\
\C_{ll} & -2.79\times 10^{-3} & -5.6\times 10^{-4} & 2.57\times 10^{-5} & 2.26\times 10^{-3} & 9.4\times 10^{-4} \\
\C_{lq}^{(3)} & 8.9\times 10^{-4} & 0 & 0 & 0 & 0 \\\end{array}
\right)
\nonumber \label{eq:fitzz2}
\eea
\noindent{Table 2: Numerical values for the coefficients defined in Eq. \ref{eq:tabdef} relevant for the on-shell process $\h\rightarrow ZZ$.}

\section{Analytical expressions of $\mathcal{F}_g$ and $\mathcal{F}_p$}
\label{sec:appf}
Here we report the explicit results for the coefficients $\mathcal{F}_g$ and $\mathcal{F}_p$ introduced in eq.~(\ref{eq:ampres}) are 

\bea
\mathcal{F}_g&=&\frac1{16 \pi ^2}\frac1{ \Lambda^2} \biggl(\frac{12 \sqrt[4]{2} \sqrt{G_\mu} \C_W M_W^3 \left(M_H^2-2 M_Z^2\right) \left(M_Z^2-6 M_W^2\right)}{M_Z^4}\nn\\
&+&\C_{\phi\square } (-6 M_H^2-12 M_t^2+9 M_W^2-\frac{2
   M_Z^2}{3})+\frac{1}{12} \C_{\phi D} \left(-9 M_H^2-36 M_t^2+8 M_W^2-35
   M_Z^2\right)\nn\\
   &+&\frac{\C_{\phi W} M_W^2 \left(M_H^2-2 M_Z^2\right) \left(9 M_H^2+18 M_t^2-56 M_W^2+3 M_Z^2\right)}{3 M_Z^4}\nn\\
      &+&\frac{\C_{\phi B} \left(M_H^2-2 M_Z^2\right)
   \left(M_Z^2-M_W^2\right) \left(9 M_H^2+18 M_t^2-100 M_W^2+85 M_Z^2\right)}{3 M_Z^4}\nn\\
   &+&\frac{\C_{\phi WB} M_W \left(M_H^2-2
   M_Z^2\right) \sqrt{M_Z^2-M_W^2} \left(3 M_H^2+18 M_t^2-42 M_W^2+56 M_Z^2\right)}{3 M_Z^4}\nn\\
   &+&\frac{2 \sqrt{2}\, \C_{uW}
   M_t M_W \left(2 M_Z^2-M_H^2\right) \left(8 M_W^2-5 M_Z^2\right)}{M_Z^4}\nn\\
   &+&\frac{2 \sqrt{2}\, \C_{uB} M_t \left(M_H^2-2 M_Z^2\right) \left(8 M_W^2-5 M_Z^2\right) \sqrt{M_Z^2-M_W^2}}{M_Z^4}\nn\\
   &+&\C_{\phi l}^{(1)} \left(8 M_Z^2-8 M_W^2\right)+\C_{\phi l}^{(3)} \left(6 M_t^2-20 M_W^2\right)+\C_{\phi e} \left(8 M_Z^2-8
   M_W^2\right)\nn\\
   &+&\C_{\phi q}^{(1)} \left(-12 M_t^2+8 M_W^2-8 M_Z^2\right)+6\,
   \C_{\phi q}^{(3)} \left(3 M_t^2-4 M_W^2\right)+4\, \C_{\phi u} \left(3 M_t^2+4 M_W^2-4 M_Z^2\right)\nn\\
   &+&\C_{\phi d} \left(8 M_Z^2-8 M_W^2\right)-3\, \C_{ll} M_Z^2-6\, \C_{lq}^{(3)} M_t^2\biggr)
   \eea
   
\bea
\mathcal{F}_p&=&\frac1{16 \pi ^2}\frac1{\Lambda ^2}\biggl(\C_{\phi\square} (-6 M_H^2-12 M_t^2+9 M_W^2-\frac{2 M_Z^2}{3})
+\frac{1}{12} \C_{\phi D} \left(-9 M_H^2-36 M_t^2+8 M_W^2-35 M_Z^2\right)\nn\\
&+&8\, \C_{\phi l}^{(1)}\left(M_Z^2-M_W^2\right)
+\C_{\phi l}^{(3)} \left(6 M_t^2-20 M_W^2\right)
+8\, \C_{\phi e} \left(M_Z^2-M_W^2\right)\nn\\
&+&\C_{\phi q}^{(1)} \left(-12 M_t^2+8 M_W^2-8 M_Z^2\right)
+6\, \C_{\phi q}^{(3)} \left(3 M_t^2-4 M_W^2\right)
+4\, \C_{\phi u} \left(3 M_t^2+4 M_W^2-4 M_Z^2\right)\nn\\
&+&8\,\C_{\phi d} \left(M_Z^2-M_W^2\right)
-3\, \C_{ll} M_Z^2-6\,\C_{lq}^{(3)} M_t^2\biggr)
\eea

\section{Analytical expression for $\Delta r$}
\label{sec:appdelta}

Here we report our result for $\Delta r$ in $R_\xi$ gauge:
\beq
16 \pi^2\Delta r=\frac{\sqrt{2} G_\mu}{2}\Delta r_{\text{SM}}+\frac1{\Lambda^2}\Delta r_{\text{EFT}},
\eeq
where $\Delta r_{\text{SM}}$ and $\Delta r_{\text{EFT}}$ are
\bea
\Delta r_{\text{SM}}&=&\left(\frac{6\, A_0\left(M_W^2\right) \left(M_H^2 \left(M_W^2-2 M_Z^2\right)+M_W^2 \left(3M_Z^2-2 M_W^2\right)\right)}{\left(M_H^2-M_W^2\right)
   \left(M_W^2-M_Z^2\right)}+4 A_0\left(M_W^2\xi\right)+\frac{6 M_W^2 \, A_0\left(M_H^2\right)}{M_H^2-M_W^2}\right.\nn\\
 &-&\left. 12 \, A_0\left(M_t^2\right)+\frac{6 \, A_0\left(M_Z^2\right) \left(2 M_W^2-
   M_Z^2\right)}{M_W^2-M_Z^2}+2 A_0\left(M_Z^2\xi\right)-M_H^2+6 M_t^2-2 M_W^2-M_Z^2\right)
\eea
and
\bea
\Delta r_{\text{EFT}}&=&\frac1{\hat{\epsilon}}\left(4\, \C_{\phi l}^{(3)} \left(3 M_t^2-4 M_W^2\right)+24\, \C_{\phi q}^{(3)} M_W^2+\frac{2}{3}\, \C_{\phi\square} M_W^2-6\, \C_{ll} M_Z^2-12 \C_{lq}^{(3)} M_t^2\right)\nn\\
&+&\Delta r_{\text{SM}} \left(\frac14 \C_{\phi D}-\C_{\phi l}^{(3)}+\frac12 \C_{ll}\right)
+\frac{6 M_W^2 \left(\text{A}_0\left(M_H^2\right)-\text{A}_0\left(M_W^2\right)\right)-M_H^4+M_W^4}{M_H^2-M_W^2}\left(\C_{\phi\square}-\frac12\C_{\phi D}\right) 
\nn\\
&+&6\left(M_t^2-2 \text{A}_0\left(M_t^2\right)\right) \left(\C_{\phi l}^{(3)}+\C_{\phi q}^{(3)}-\C_{lq}^{(3)}-\frac14\C_{\phi D}\right)
+6 \text{A}_0\left(M_Z^2\right) \left(\C_{ll}+c_W^2\left(4 C_{\phi l}^{(3)}-\C_{\phi D} \right)\right)
\nn\\
&-&12 M_Z^2 \left(c_W s_W \C_{\phi WB}-\frac{5}{3} c_W^2 \C_{\phi l}^{(3)}-\frac{5}{12} \C_{ll} - \frac16 \C_{\phi D}\right)
\nn\\
&+&12\left(\text{A}_0\left(M_W^2\right)-c_W^2\text{A}_0\left(M_Z^2\right)\right)\left(\C_{ll}+\C_{\phi l}^{(3)}+\C_{\phi l}^{(1)}\right)
\nn\\
&-& \left(\text{A}_0\left(\xi  M_W^2\right)+\text{A}_0\left(\xi  M_Z^2\right)+2 \left(M_W^2+M_Z^2\right)\right) \C_{\phi D},
\eea
and as usual $c_W=\frac{M_W^2}{M_Z^2}$, $s_W=\sqrt{1-c_W^2}$ and
\beq
A_0(x)=-i\frac{(2 \pi \Lambda)^{2\epsilon}}{\pi^2}\int\frac{d^d k_1 }{(k_1^2-x)}=\frac{x}{\hat{\epsilon}}+ x(1-\log(\frac{x}{\Lambda^2}))
\eeq

\bibliographystyle{utphys}
\bibliography{hzz_v4}

\end{document}